\documentclass[12pt, journal, onecolumn]{IEEEtran}
\date{}

\usepackage{subfigure}
\usepackage{amsmath,amssymb}
\usepackage{color}
\usepackage{graphicx}
\usepackage{setspace}

\newtheorem{theorem}{Theorem}
\newtheorem{proposition}{Proposition}
\newtheorem{lemma}{Lemma}
\newtheorem{corollary}{Corollary}
\newtheorem{remark}{Remark}
\allowdisplaybreaks

\begin{document}

\title{Iterative Dynamic Water-filling for Fading Multiple-Access Channels with Energy Harvesting}

\author{Zhe~Wang,  Vaneet~Aggarwal,~\IEEEmembership{Member,~IEEE} and~Xiaodong~Wang,~\IEEEmembership{Fellow,~IEEE}
\thanks{Z. Wang and X. Wang are with the Electrical Engineering Department, Columbia University, New York, NY 10027 (e-mail: \{zhewang, wangx\}@ee.columbia.edu).

V. Aggarwal is with AT\&T Labs-Research, Florham Park, NJ 07932 USA (e-mail: vaneet@research.att.com).
}}

\maketitle

\begin{abstract}
In this paper, we develop optimal energy scheduling algorithms for $N$-user fading multiple-access channels with energy harvesting to maximize the channel sum-rate, assuming that the side information of both the channel states and energy harvesting states for $K$ time slots is known {\em a priori}, and the battery capacity and the maximum energy consumption in each time slot are bounded. The problem is formulated as a convex optimization problem with ${\cal O}(NK)$ constraints making it hard to solve using a general convex solver since the computational complexity of a generic convex solver is exponential in the number of constraints. This paper gives an efficient energy scheduling algorithm, called the iterative dynamic water-filling algorithm, that has a computational complexity of ${\cal O}(NK^2)$ per iteration. For the single-user case, a dynamic water-filling method is shown to be optimal. Unlike the traditional water-filling algorithm, in dynamic water-filling, the water level is not constant but changes when the battery overflows or depletes. An iterative version of the dynamic water-filling algorithm is shown to be optimal for the case of multiple users. Even though in principle the optimality is achieved under large number of iterations, in practice convergence is reached in only a few iterations. Moreover, a single iteration of the dynamic water-filling algorithm achieves a sum-rate that is within $(N-1)K/2$ nats of the optimal sum-rate.
\end{abstract}


\section{Introduction}

Energy harvesting has grown from long-established concepts into devices for powering ubiquitously deployed sensor networks and mobile electronics. This is because the use of energy harvesting devices can prolong the lifetime of a transmitter (or a user)~\cite{Energy_Scav}\cite{EHSNSI}. On the other hand, many challenging research issues arise from the new paradigm of communication powered by harvested energy. In particular, the maximum achievable rate under the dynamic energy constraints is one fundamental problem, especially for the case of fading multiple-access  channels (MAC)~\cite{OPSMAC}. This paper addresses this by developing an optimal energy scheduling algorithm for a fading MAC.

For the single-user case, the traditional water-filling algorithm provides the optimal power control for fading channels with an average power constraint~\cite{IT}. Single-user channels with an energy harvesting device have also been studied.  In a static channel with finite battery capacity, the optimality of the shortest-path-based algorithm is discussed in \cite{FHEARS} for the energy scheduling. The authors of \cite{OTPBLE} analyzed the optimality properties based on the energy causality and provide an algorithm to obtain the energy schedule. In \cite{OPSEHC}, the optimal utilization of the harvested energy was considered to optimize packet transmission time. In a broadcast channel with energy harvesting, the optimal energy scheduling was discussed in \cite{OBSEHR} using directional water-filling. For dynamic fading channels with non-causal side information, the optimal energy allocation with energy harvesting constraints was treated in \cite{OEAWCE}, and a staircase water-filling algorithm is proposed for the case of infinite battery capacity. Taking account into the finite battery capacity, the energy-flow behavior with an energy harvesting device was discussed in \cite{TEHNFW} and the method of the directional water-filling was proposed. Moreover, the authors of \cite{TMGRCE} discussed the scheduling problem in the Gaussian relay channel with energy harvesting. However, to the best of our knowledge, so far no study considers the maximum per-slot energy consumption in a fading channel with a finite battery capacity for the energy harvesting transmitter.  In this paper we develop an optimal energy scheduling algorithm, called the dynamic water-filling algorithm, for fading channels with both battery capacity and  maximum per-slot energy consumption constraints, to maximize the sum of the channel rate.

In particular, we first consider the energy scheduling for a single-user fading channel with energy harvesting, constrained by the availability of the energy, the capacity of the battery, and the maximum per-slot energy consumption of the transmitter. We assume that both the channel state and the energy harvesting state are non-causally known and we use the sum of the channel rate over $K$ time slots as the performance metric of the energy scheduler. We propose a dynamic water-filling algorithm that consists of three phases. In the first phase, we calculate the optimal energy wastage schedule. Using this energy wastage schedule, in the second phase, we calculate the optimal {\em battery depletion points}  (BDPs) and {\em battery fully-charged points} (BFPs) that represent the time slots where the water levels change. In the third phase,  we apply the water-filling method to each segment between adjacent optimal BDPs and/or BFPs.  The optimality of this dynamic water-filling algorithm is shown. Moreover, when the battery capacity and the transmission power are unconstrained, the proposed dynamic water-filling algorithm reduces to the staircase water-filling method given in \cite{OEAWCE}, with non-decreasing water levels over time slots.

The fading MAC has been widely studied with an average power constraint \cite{IWGVMC}\cite{MFCPSO}\cite{ICPCSM}. An iterative water-filling algorithm was proposed in \cite{IWGVMC} for Gaussian vector fading MAC with average power constraints. From the viewpoint of packet transmission with energy harvesting, \cite{OPSMAC} adapted an optimal iterative directional water-filling algorithm and obtained the general capacity region for a static MAC without finite battery capacity and maximum power (per-slot energy consumption) constraints. In this paper, we formulate a convex optimization problem with ${\cal O}(NK)$ constraints for maximum-sum-rate energy scheduling in a fading MAC with finite battery capacity and maximum per-slot energy consumption constraints. Since the computational complexity of a generic convex solver is exponential in the number of constraints, we need to develop an efficient algorithm to obtain the optimal energy schedule.

To that end, we propose an iterative dynamic water-filling algorithm, that is an iterative version of the dynamic water-filling algorithm for single-user channels. The per-iteration complexity of the iterative dynamic water-filling algorithm is ${\cal O}(NK^2)$, which is much lower than that of a general convex solver. Also, we show that a single iteration of the dynamic water-filling algorithm achieves a sum-rate that is within $(N-1)K/2$ nats of the optimal sum-rate. Moreover, simulations show that the iterative dynamic water-filling algorithm converges within only a few iterations.

The remainder of the paper is organized as follows. In Section II, we describe the system model and formulate the energy scheduling problem as a convex optimization problem. In Section III, we treat the single-user case and develop the dynamic water-filling algorithm as the optimal energy scheduler. In Section IV, we treat the multi-user case and develop the iterative dynamic water-filling algorithm. Simulation results are provided in Section V. Finally, Section VI concludes the paper.

\section{Problem Formulation}
\begin{figure}[!hbp]
\centering
\includegraphics[width=.9\textwidth]{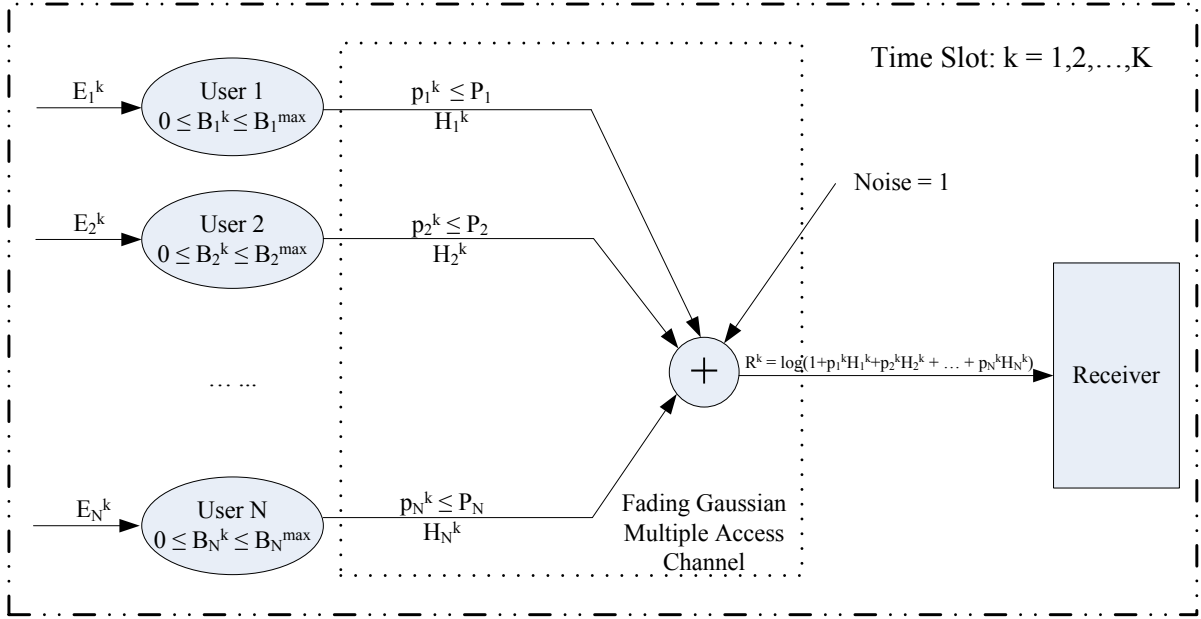}
\caption{The system diagram.}\label{fg:sys}
\end{figure}

We consider a fading multiple-access channel with $N$ transmitters and one receiver as shown in Fig.~\ref{fg:sys}. We assume a time-slotted model with $K$ time slots. Each slot consists of $T$ time instants during which the channel from each transmitter to the receiver remains constant. Transmitter $n$ transmits signal $X_{nki}$ at instant $i$ in slot $k$, and the signal received at the receiver at instant $i$ in slot $k$ is given by $Y_{ki} =\frac{1}{\sqrt{T}} \sum_{n=1}^N h_{nk}X_{nki}+Z_{ki}$ where and $h_{nk}$ is the complex channel gain from transmitter $n$ to the receiver in slot $k$ and $Z_{ki}\sim \mathsf{CN}(0,1)$ is the i.i.d. complex Gaussian noise sample.  We denote ${\cal N}\triangleq \{1,2,\ldots, N\}$ as the set of transmitters,  ${\cal K} \triangleq  \{1,2,\ldots,K\}$ as the set of the time slots, and  $p_n^k$ as the energy consumed by transmitter $n$ in slot $k$ ($p_n^k \triangleq\frac{1}{T} \sum_{i=1}^T |X_{nki}|^2$). Denote $H_n^k \triangleq |h_{nk}|^2$. The upper bound on the reliable transmission rate of the MAC over $K$ slots is given by~\cite{IT},
\begin{equation}\label{eq:obj}
C({\cal P}) \triangleq \sum_{k\in{\cal K}}\log(1+\sum_{n\in{\cal N}}p_n^kH_n^k)
\end{equation}
where ${\cal P} \triangleq  \{\boldsymbol{p}_n\;|\; \boldsymbol{p}_n \triangleq [p_n^1,p_n^2,\ldots,p_n^K], \ n\in {\cal N}\}$. We note that this upper bound can be achieved by transmitting i.i.d. Gaussian signals as $T\rightarrow \infty$.

We assume that each transmitter is equipped with an energy harvesting device and a buffer battery. The energy harvesting device, e.g., the solar panel, harvests energy from the surrounding environment. We denote the total energy harvested up to the end of the $k$-th slot for transmitter $n$ as $E_n^k$.  Since the energy harvesting may be practically predicted with the state-of-art techniques \cite{PMEHW}\cite{AMPEAE}, especially with good accuracy for the short-term prediction,  we assume that the amount of the harvested energy in each slot is perfectly estimated in our model. For channel fading, the current techniques make the short-term prediction also possible \cite{FCPMRA}\cite{AMLPMF}, especially for the slow fading channel considered in our model. Therefore, we consider that $H_n^k,\ E_n^k,\ n\in{\cal N},\ k\in{\cal K}$ are known non-causally to all transmitters.

Denote $B_n^k$ as the battery energy level in slot $k$ at transmitter $n$.  We further assume that each battery has a limited capacity, denoted by $B_n^{\max}$, i.e., $B_n^k \leq B_n^{\max}$, and each transmitter has a maximum energy consumption constraint that the scheduled energy consumption in each slot cannot exceed certain value $P_n$, i.e., $p_n^k\leq P_n$, $k\in{\cal K},n\in{\cal N}$. Moreover, we note that $p_n^k\geq 0$ and $B_n^k \geq 0$, for all $k\in{\cal K},n\in{\cal N}$. Without loss of generality, we can assume that the battery is empty initially. Then the battery level at the end of slot $k$ is given by
\begin{equation}\label{eq:battery}
B_n^{k}  \triangleq \min\big\{B_n^{\max}, E_n^k - \sum_{t=1}^k p_n^{t}\big\}\ .
\end{equation}

With the maximum energy consumption constraint and the finite battery capacity, we may not be able to utilize all harvested energy or even store the unused energy, i.e., even if we use the maximum energy assumption to transmit in each slot, the accumulated surplus energy may still tend to exceed the battery capacity in some slots. In order to meet the maximum battery capacity constraint, some harvested energy needs to be wasted. Denote $D_n^k \geq 0$ as the amount of energy wasted at transmitter $n$ in slot $k$, such that
\begin{equation}\label{eq:dischage}
E_n^k - \sum_{t=1}^k p_n^{t} - \sum_{t=1}^k D_n^{t}\leq B_n^{\max}\ .
\end{equation}
Then, the battery level in \eqref{eq:battery} can be rewritten as
\begin{equation}  \label{eq:lbattery}
B_n^{k} =  E_n^k - \sum_{t=1}^k p_n^{t} - \sum_{t=1}^k D_n^{t}\ .
\end{equation}

Denote the {\em wastage schedule} ${\cal D} \triangleq  \{\boldsymbol{D}_n\;|\; \boldsymbol{D}_n \triangleq [D_n^1,D_n^2,\ldots,D_n^K], \ n\in {\cal N}\}$. 
Then we formulate the following optimization problem of maximizing the sum-rate over  {\em energy schedule}  consisting of the {\em transmission schedule} $\cal P$ and the wastage schedule $\cal D$:
\begin{equation}\label{eq:problem}
\max_{{\cal P,D} } C({\cal P})
\end{equation}
subject to
\begin{equation}\label{eq:cst}
\left\{\begin{array}{l}
0 \leq  E_n^k - \sum_{t=1}^k p_n^{t} - \sum_{t=1}^k D_n^{t} \leq B_n^{\max} \\
0 \leq p_n^k \leq P_n\\
D_n^k \geq 0
\end{array}
\right. ,
\end{equation}
for all $k \in {\cal K},\ n\in{\cal N}$.

Note that this problem is a convex optimization problem and can be solved by a general convex solver, whose complexity is however exponential in the number of constraints~\cite{CO}, which in this case is ${\cal O}(NK)$. To reduce the computational complexity, we will develop efficient solutions in this paper, by exploiting the specific properties of the optimal energy schedule that we will identify.

\section{Dynamic Water-filling for Single-user Case}

In this section, we give an optimal solution to the energy scheduling problem in  \eqref{eq:problem} for the single-user case with ${\cal N}=\{1\}$. For simplicity, we will drop the subscripts in all terms. Denoting
$C_s(\boldsymbol{p}) \triangleq \sum_{k\in{\cal K}}\log(1+p^kH^k)$, the single-user version of the problem in \eqref{eq:problem}-\eqref{eq:cst} is rewritten as
\begin{equation}\label{eq:sproblem}
\max_{\boldsymbol{p},\boldsymbol{D}} C_s(\boldsymbol{p})
\end{equation}
subject to
\begin{equation}\label{eq:scst}
\left\{\begin{array}{l}
0\leq E^k - \sum_{t=1}^{k}p^{t}-\sum_{t=1}^{k}D^{t}  \leq B^{\max}\\
0 \leq p^k \leq P\\
D^k \geq 0,
\end{array}
\right.
\end{equation}
for all $k \in {\cal K}$.

Given a wastage schedule $\boldsymbol{D}$, we rewrite the problem in \eqref{eq:sproblem}-\eqref{eq:scst} as
\begin{equation}\label{eq:opts}
S(\boldsymbol{D}) \triangleq \max_{\boldsymbol{p}}C_s(\boldsymbol{p})
\end{equation}
subject to:
\begin{equation}\label{eq:optscst}
\left\{\begin{array}{l}
 E^k -  B^{\max} - \sum_{t=1}^{k}D^{t} \leq \sum_{t=1}^{k}p^{t} \leq E^k - \sum_{t=1}^{k}D^{t}\\
0 \leq p^k \leq P\\
\end{array}
\right.
\end{equation}for all $k \in {\cal K}$. Then we solve the single-user problem with two successive stages. In the first stage, we obtain the optimal wastage schedule $\boldsymbol{D}^*$, such that
\begin{equation}\label{eq:stage1}
S(\boldsymbol{D}^*) = \max_{\boldsymbol{p},\boldsymbol{D}} C_s(\boldsymbol{p})
\end{equation}
subject to the constraints in \eqref{eq:scst} for all $k\in{\cal K}$. In the second stage, applying the optimal wastage schedule $\boldsymbol{D}^*$, we then obtain the optimal transmission schedule $\boldsymbol{p}^*$ for the problem $S(\boldsymbol{D}^*)$ subjecting to the constraints in \eqref{eq:optscst} for all $k \in {\cal K}$, i.e, finally the obtained transmission schedule $\boldsymbol{p}^*$ and $\boldsymbol{D}^*$ is the optimal solution to the problem in \eqref{eq:sproblem}-\eqref{eq:scst}.

\subsection{The Optimal Wastage Schedule}
Based on the battery level dynamic in \eqref{eq:lbattery}, the harvested energy may be wasted if the transmitter cannot use all available energy due to the maximum energy consumption constraint and the surplus energy cannot be deposited into the battery due to the battery capacity constraint. To obtain the optimal wastage schedule $\boldsymbol{D}^*$, we consider a scenario where all energy waste is due to energy overflow, i.e.,
\begin{align}
{D^*}^k& = \max\big\{B^{k-1} + (E^k - E^{k-1}) - p^k - B^{\max},0\big\}\label{eq:greedy}
\end{align}
where  $p^k$ is chosen by the following the greedy policy
\begin{equation}\label{eq:greedyp}
p^k = \min\big\{P, B^{k-1} + E^k - E^{k-1}\big\}\ .
\end{equation}

In the above schedule, ${D^*}^k$ is determined by $B^{k-1}$ only given all $E^k$ are known. Since it is assumed that $B^0=0$,  ${D^*}^k$ can be obtained by computing \eqref{eq:greedyp}, \eqref{eq:greedy} and \eqref{eq:lbattery} for $k=1,2,\ldots,K$, shown in the following table.\\
\begin{minipage}[h]{6.5 in}
\rule{\linewidth}{0.3mm}\vspace{-.05in}
{\bf {\footnotesize Algorithm 1 - Algorithm for computing the optimal wastage schedule}}\vspace{-.1in}\\
\rule{\linewidth}{0.2mm}
{ {\small
\begin{tabular}{ll}
    \;& $B^0=0$\\
	\;&\textbf{\bf FOR} $k=1,2,\ldots,K$\\
        \;&\quad Obtain $p^k$ based on $B^{k-1}$ by \eqref{eq:greedyp}\\
        \;&\quad Obtain ${D^*}^k$ based on $p^k$ by \eqref{eq:greedy}\\
        \;&\quad Obtain $B^k$ based on ${D^*}^k$ and $p^k$ by \eqref{eq:lbattery}\\
	\;&\textbf{\bf ENDFOR}\\
\end{tabular}}}\\
\rule{\linewidth}{0.3mm}
\end{minipage}\vspace{.002 in}\\

Note that, according to the procedure to obtain $\boldsymbol{D}^*$, there always exists feasible transmission schedule $\boldsymbol{p}$ such that $\boldsymbol{D}^*$ and $\boldsymbol{p}$ satisfy the constraints in \eqref{eq:scst}, i.e., the feasible domain in \eqref{eq:scst} is non-empty under $\boldsymbol{D}^*$. Moreover, we say a wastage schedule $\boldsymbol{D}$ is {\em feasible} if and only if there is a feasible value of $p^k$ for this wastage schedule. For example, $\boldsymbol{D}^*$ is a feasible wastage schedule.

For any feasible wastage schedule $\boldsymbol{D}$, the following lemma indicates that it wastes no less total energy than $\boldsymbol{D}^*$.
\begin{lemma}\label{lm:1}
For any energy wastage schedule $\boldsymbol{D}=[D^1,D^2,\ldots,D^K]$, if $\sum_{k\in{\cal K}}D^k<\sum_{k\in{\cal K}}{D^*}^k$, then the feasible domain in \eqref{eq:scst} under $\boldsymbol{D}$ is empty, i.e., $\boldsymbol{D}$ is infeasible.
\end{lemma}
\begin{IEEEproof}
Since with $\boldsymbol{D}^*$, the maximum possible amount of the harvested energy is used for transmission in each slot, the amount of the wasted energy is minimized subject to a non-empty feasible domain in \eqref{eq:scst}. Obviously, for any other energy wastage schedule $\boldsymbol{D}$ that wastes less energy than $\boldsymbol{D}^*$, the feasible domain in \eqref{eq:scst} under $\boldsymbol{D}$ must be empty.
\end{IEEEproof}

Moreover, the next lemma shows that under any feasible $\boldsymbol{D}$ with the same total energy wastage as $\boldsymbol{D}^*$, the optimal values for the optimization problem in \eqref{eq:sproblem}-\eqref{eq:scst} are same. The proof is given in Appendix A.
\begin{lemma}\label{lm:2}
For any feasible energy wastage schedule $\boldsymbol{D}=[D^1,D^2,\ldots,D^K]$ such that $\sum_{k\in{\cal K}}D^k=\sum_{k\in{\cal K}}{D^*}^k$,  the optimal values of \eqref{eq:sproblem} under $\boldsymbol{D}$ and $\boldsymbol{D}^*$ are  same.
\end{lemma}

We next show in the following theorem that the optimal transmission schedule for the problem in \eqref{eq:sproblem}-\eqref{eq:scst} is obtained by solving problem $S(\boldsymbol{D}^*)$,  i.e., the equality in \eqref{eq:stage1} holds.

\begin{theorem}\label{thm:oo}
The optimal values of the problems in \eqref{eq:sproblem}-\eqref{eq:scst} and in \eqref{eq:opts}-\eqref{eq:optscst} with $\boldsymbol{D} =\boldsymbol{D}^*$ are the same, i.e.,
\begin{equation}
S(\boldsymbol{D}^*) = \max_{\boldsymbol{p},\boldsymbol{D}}C_s(\boldsymbol{p})
 \end{equation}
subject to the constraints in \eqref{eq:scst} for all $k\in{\cal K}$.
\end{theorem}
\begin{IEEEproof} We first note by Lemma \ref{lm:1} that for any energy wastage schedule $\boldsymbol{D}=[D^1,D^2,\ldots,D^K]$ such that $\sum_{k\in{\cal K}}D^k<\sum_{k\in{\cal K}}{D^*}^k$, there is no feasible solution to problem $S(\boldsymbol{D})$.

According to the causality of the energy harvesting, i.e., the harvested energy stored in the battery can be consumed and/or wasted in any slot as long as the corresponding energy is still stored in the battery, any feasible schedule $\boldsymbol{D}$ such that $\sum_{k\in{\cal K}}D^k>\sum_{k\in{\cal K}}{D^*}^k$ can be generated from some feasible $\tilde{\boldsymbol{D}}=[\tilde D^1,\tilde D^2,\ldots,\tilde D^K]$ such that $\sum_{k\in{\cal K}}{\tilde D}^k = \sum_{k\in{\cal K}}{D^*}^k$ by increasing some of ${\tilde D}^k$ to $D^k$ under the constraints in \eqref{eq:scst}.

For any wastage schedule $\boldsymbol{D}$ such that $\sum_{k\in{\cal K}}D^k>\sum_{k\in{\cal K}}{D^*}^k$, we find the optimal $\boldsymbol{p}$ optimizing $S(\boldsymbol{D})$. By decreasing the energy wastage, the energy allocation in each slot can be increased by $D^k - {\tilde D}^k $ to give a new energy allocation $\tilde{\boldsymbol{p}}$ such that $p^k \leq {\tilde p}^k$ and $\sum_{k\in{\cal K}}{\tilde D}^k=\sum_{k\in{\cal K}}{D^*}^k$. Since $p^k \leq {\tilde{p}}^k$, we have $S(\boldsymbol{D})\leq S(\tilde{\boldsymbol{D}})$. By Lemma \ref{lm:2}, we have $S(\tilde{\boldsymbol{D}})=S(\boldsymbol{D}^*)$. Thus, $S(\boldsymbol{D})\leq S(\boldsymbol{D}^*)$ for any feasible $\boldsymbol{D}$, thus showing the optimality of $\boldsymbol{D}^*$ and $S(\boldsymbol{D}^*)$.
\end{IEEEproof}

\begin{figure}[!hbp]
\centering
\includegraphics[width=.9\textwidth]{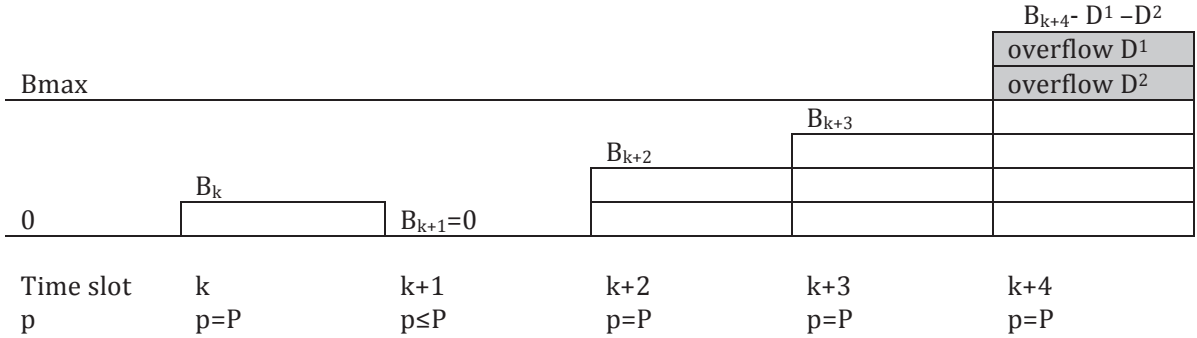}
\caption{An example of ${D^*}$. Following Algorithm 1, in each slots the transmitter intends to use as much energy as possible. Specifically, since battery state in slot $k,k+2,k+3,k+4$ is non-empty, the transmission energy in these slots must achieve the maximum energy consumption constraint. Moreover, in slot $k+4$, since the remaining energy tends to exceed the battery capacity, the amount of $D^1+D^2$ energy needs to be overflowed.}\label{fg:d1}
\end{figure}

\begin{figure}[!hbp]
\centering
\includegraphics[width=.9\textwidth]{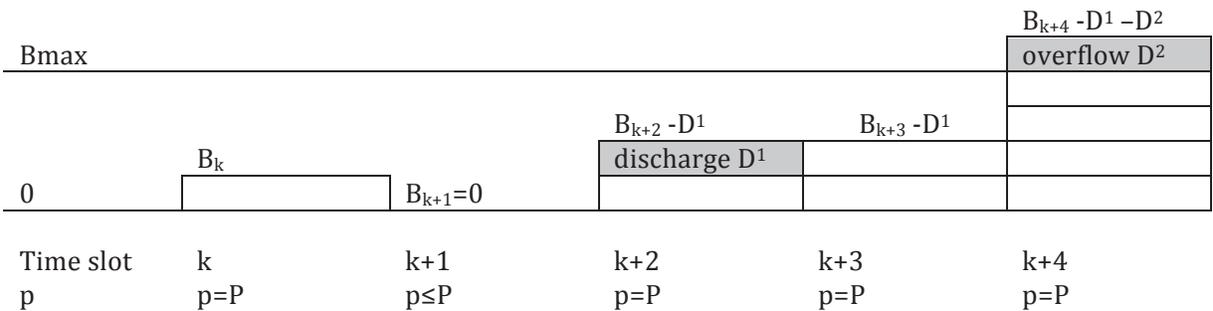}
\caption{An example of rescheduling ${D^*}$ in Fig. \ref{fg:d1}. For $D^*$, the amount of $D^1+D^2$ energy needs to be overflowed in slot $k+4$. Without violating the constraints in \eqref{eq:scst}, the amount of $D^1$ energy can be wasted in slot $k+2$, instead of overflowing in slot $k+4$. With this change, $p^k$ is not changed but the overflowed energy in slot $k+4$ is reduced to $D^2$.}\label{fg:d2}
\end{figure}

\begin{remark} \label{rm:1}{Algorithm 1 provides a method to calculate the optimal energy wastage schedule ${\cal D}^*$. However, by Lemma \ref{lm:2}, it is easy to see that the optimal energy wastage schedule ${\cal D}$ may not be unique. Then, according to the proof of Theorem \ref{thm:oo}, we can further have that any energy wastage schedule $D^k$ that is feasible and satisfies $\sum_{k\in{\cal K}}D^k=\sum_{k\in{\cal K}}{D^*}^k$ is optimal. Even though the optimal wastage schedule is not unique, we have that the domain for the transmission schedule for all these optimal values of wastage schedule is the same and thus the optimal value of transmission schedule is unique. An example is shown in Figures \ref{fg:d1} and \ref{fg:d2}, where two different optimal wastage schedules lead to the same feasible domain for the transmission schedule. }
\end{remark}

{Specifically, Fig.~\ref{fg:d1} shows the energy wastage schedule and the corresponding transmission schedule obtained by Algorithm 1. In the slots with non-empty battery, e.g., $k+1$, $k+2$, $k+3$ and $k+4$, its corresponding transmission schedule is $P$ according to \eqref{eq:greedyp}. In the slot that the battery level tends to exceed $B^{\max}$, e.g., $k+4$, the exceeded energy has to be wasted, say $D^1+D^2$. To generate another wastage schedule $\boldsymbol{D}$ such that $\sum_{k\in{\cal K}}D^k= \sum_{k\in{\cal K}}{D^*}^k$, we can reschedule part of the wasted energy, e.g., $D^1$ in slot $k+4$, to be wasted in another slot, e.g., $k+2$, as illustrated in Fig.~\ref{fg:d2}. Since the wasted energy, e.g., $D^1+D^2$ in slot $k+4$, cannot be rescheduled to be wasted in the slot before the energy is harvested, e.g., $k$ and $k+1$, the transmission schedule in the originally scheduled slot, e.g., $k+4$, and the rescheduled slot, e.g., $k+2$, should remain $P$, otherwise the feasible domain in \eqref{eq:scst} becomes empty. Therefore, given ${\boldsymbol{D}^*}$ and $\boldsymbol{D}$ such that $\sum_{k\in{\cal K}}D^k=\sum_{k\in{\cal K}}{D^*}^k$, the feasible domains of $\boldsymbol{p}$ are exactly the same. Therefore, the optimal transmission schedule is exactly the same.}


\subsection{Properties of Optimal Transmission Schedule}
By Theorem \ref{thm:oo}, we know that the optimal transmission schedule $\boldsymbol{p}^*$ can be obtained by solving the problem $S(\boldsymbol{D}^*)$, where $\boldsymbol{D}^*$ is the optimal wastage schedule obtained by Algorithm 1. In this subsection, we describe the optimality conditions for the optimal transmission schedule $\boldsymbol{p}^*$.

Denoting ${\tilde E}^k\triangleq E^k  - \sum_{t=1}^k {D^*}^{t}$, we rewrite the problem $S(\boldsymbol{D}^*)$ as
\begin{equation}\label{eq:nsp}
 \max_{\boldsymbol{p}}C_s(\boldsymbol{p}),
\end{equation}
subject to
\begin{equation}\label{eq:nspcst}
\left\{\begin{array}{l}
0 \leq {\tilde E}^k - \sum_{t=1}^{k}p^{t} \leq B^{\max}\\
0 \leq p^k \leq P\\
\end{array}
\right.\ ,
\end{equation}
for all $k\in{\cal K}$.

{In the above problem, ${\tilde E}^k$ is the ``effective energy'' after taking the energy wastage into account. This problem would utilize all this effective energy and there will be no additional wastage. This is because any wastage can be avoided by using it as the transmission energy in some time slot and thus would result in a higher objective value.}

The Lagrangian function for the problem in \eqref{eq:nsp}-\eqref{eq:nspcst} with auxiliary variables  $\lambda^k \geq 0$, $\mu^k \geq 0$, $\alpha^k\geq0$, and $\beta^k \geq 0$ is given by
\begin{align}
{\cal L}\triangleq\;\; C_s(\boldsymbol{p})& - \sum_{k\in{\cal K}}(p^k-\tilde{E}^k) \sum_{t=k}^K\lambda^t  \nonumber\\&+ \sum_{k\in{\cal K}}(p^k-\tilde{E}^k + B^{\max})\sum_{t=k}^K\mu^{t}\nonumber\\
 &- \sum_{k\in{\cal K}}\alpha^k(p^k-P)  + \sum_{k\in{\cal K}}\beta^kp^k\end{align}

Note that the objective function in \eqref{eq:nsp} is the sum of concave functions and the constraints in \eqref{eq:nspcst} are linear. Hence, the Karush-Kuhn-Tucker (KKT) conditions are necessary and sufficient conditions for optimality~\cite{CO}, which are
\begin{align}
\frac{H^k}{1+p^kH^k} = v^k - u^k + \alpha^k - \beta^k,\ k\in{\cal K}\label{kkt:wf}\\
\lambda^k\cdot(\sum_{t=1}^kp^{t}-{\tilde E}^k)=0,\ k\in{\cal K}\label{kkt:a}\\
\mu^k\cdot (\sum_{t=1}^kp^{t}-{\tilde E}^k+B^{\max})=0,\ k\in{\cal K}\label{kkt:b}\\
\alpha^k\cdot (p^k-P) = 0,\ k\in{\cal K}\label{kkt:c}\\
\beta^k\cdot p^k=0,\ k\in{\cal K}\label{kkt:d}
\end{align}
as well as the constraints in \eqref{eq:scst} and $\lambda^k,\mu^k,\alpha^k,\beta^k\geq 0,\ k\in{\cal K}$,
where
\begin{equation}
u^k \triangleq \sum_{t=k}^K\mu^{t},\ v^k \triangleq \sum_{t=k}^K\lambda^{t}\label{eq:l2}\ .
\end{equation}

Then, we can obtain the necessary and sufficient conditions for the optimality of the transmission schedule $\boldsymbol{p}^*$, satisfying the water-filling strategy, where the water levels change at some specific slots.

\begin{theorem}\label{thm:wf}
A transmission schedule $\boldsymbol{p}$ is the optimal solution to \eqref{eq:nsp}-\eqref{eq:nspcst}, if and only if,
\begin{equation}\label{eq:waterfilling}
p^k = \min\Big(P\;,\;\left[\frac{1}{v^k-u^k} - \frac{1}{H^k}\right]^+\Big)
\end{equation}
which is a form of water-filling solution where the water level $\frac{1}{v^k-u^k}$ may only increase at the battery depletion points (BDPs), i.e., $k\in {\cal K}$, such that $B^k =0$, and decrease at the battery fully-charged points (BFPs), i.e., $k\in {\cal K}$, such that $B^k=B^{\max}$, and the resulted solution is feasible.
\end{theorem}
\begin{IEEEproof}
The first KKT condition in \eqref{kkt:wf} can be rewritten as
\begin{equation}
p^k = \frac{1}{v^k-u^k+\alpha^k-\beta^k}- \frac{1}{H^k}\ .
\end{equation}
Thus, the optimal energy level in slot $k$ is determined by the dual variables $v^k, u^k, \alpha^k, \beta^k$ and the channel state $H^k$. From \eqref{kkt:c} and \eqref{kkt:d}, we see that if $p^k<P$, $\alpha^k=0$ and if $p^k>0$, $\beta^k=0$ and thus we have \eqref{eq:waterfilling}.

The KKT conditions in \eqref{kkt:a} and \eqref{kkt:b} constrain the changes of the water levels $\frac{1}{v^k-u^k}$. To satisfy these two conditions, $\lambda^k$ may only be non-zeros at the BDPs, i.e., for $k$ such that ${\tilde E}^k-\sum_{t=1}^k p^{t}=B^k =0$, and $\mu^k$ may only be non-zeros at the BFPs, i.e., for $k$ such that ${\tilde E}^k-\sum_{t=1}^kp^{t}=B^k = B^{\max}$. Since the dual variables $\lambda^k$ and $\mu^k$ are non-negative, $u^k$ and $v^k$, which are defined in \eqref{eq:l2}, are non-increasing over $k$. Specifically, since $\lambda^k$ and $\mu^k$ cannot be both non-zero at the same time, the water level $\frac{1}{v^k-u^k}$ may only increase when $v^k$ changes, i.e., $\lambda^k$ is non-zero at the BDPs, or decrease when  $u^k$ changes, i.e., $\mu^k$ is non-zero at the BFPs.
\end{IEEEproof}

Theorem \ref{thm:wf} shows that the optimal transmission schedule must satisfy a dynamic level water-filling condition, as well as the resulted transmission schedule must be feasible. Specifically, for water-filling, we apply different water levels on some consecutive time slots, namely {\em segment}; and the water level may only increase at the BDPs, i.e., the slot when the resulted transmission schedule causes the battery empty, and decrease at the BFPs, i.e., the slot when the resulted transmission schedule causes the battery fully-charged.

\begin{figure}[!hbp]
\centering
\includegraphics[width=.9\textwidth]{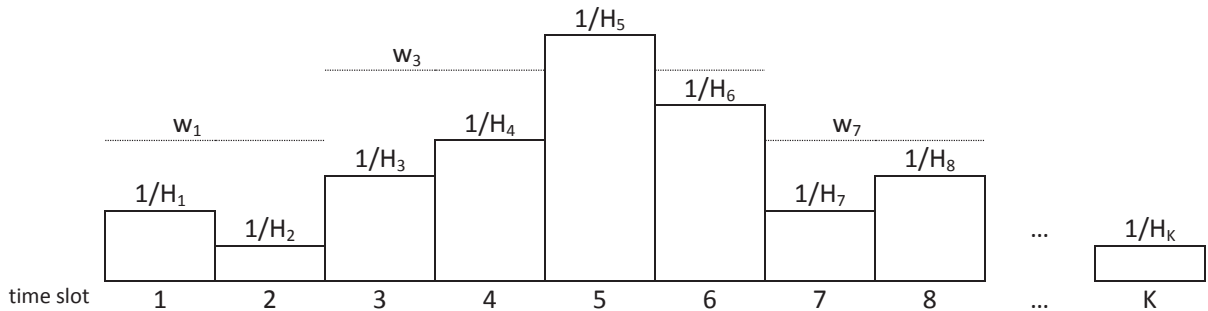}
\caption{An example of a transmission schedule and the corresponding water levels and BDPs/BFPs.}\label{fg:rs}
\end{figure}

For example, given any water levels and their corresponding segments, the relationship among the water level $w^k$, the transmission schedule $p^k$, and the channel state $H^k$ can be characterized as in Fig.~\ref{fg:rs}. In particular, if the transmission schedule shown in Fig.~\ref{fg:rs} is optimal for the problem in \eqref{eq:nsp}-\eqref{eq:nspcst}, the water level $w^k=1/(v^k-u^k)$ can only increase at the BDPs, e.g., slot $2$ , and decrease at the BFPs, e.g., slot $6$, as well as the resulted transmission schedule is feasible, i.e., the constraints in \eqref{eq:nspcst} are satisfied.

Given the optimal transmission schedule, by Theorem \ref{thm:wf}, we can obtain a set of the resulted BDP/BFPs, denoting as the {\em optimal BDP/BFPs set} ${\cal X}^*$. Specifically, ${\cal X}^*$  must contain $(0,\textrm{BDP})$ and $(K,\textrm{BDP})$ by default. Assuming the elements in ${\cal X}^*$ are sorted by ascending order of the associated slot, then given two adjacent BDP/BFP in ${\cal X}^*$, e.g., $(a,\textrm{type of } a)$ and $(b,\textrm{type of } b)$, by Theorem \ref{thm:wf}, we have the one-to-one mapping to the optimal transmission schedule, and the optimal water level:
\begin{equation}\label{eq:swfc}
p^k = \min\Big(P\;,\;\Big[\frac{1}{w} - \frac{1}{H^k}\Big]^+\Big)\ ,
\end{equation}
where $w$ is the {\em water level of a segment} such that
\begin{align}
&\sum_{t=a+1}^b p^{t} = \nonumber \\
&\min\Big\{(b-a)P,{\tilde E}^b-{\tilde E}^{a} + \big(\mathbb{I}(a\textrm{ is BFP})-\mathbb{I}(b\textrm{ is BFP})\big)B^{\max}\Big\}\ ,\label{eq:waterlevel}
\end{align}
with $\mathbb{I}(\cal A)$ being an indicator function given by
\begin{equation}
\mathbb{I}({\cal A})\triangleq
\left\{\begin{array}{ll}
1,&\textrm{if } {\cal A} \textrm{ is true}\\
0,&\textrm{otherwise}\\
\end{array}\right.\ .
\end{equation}
Note that, \eqref{eq:swfc}-\eqref{eq:waterlevel} represent the water-filling operation in a segment between two BDP/BFPs, as mentioned in Theorem \ref{thm:wf}. Also, \eqref{eq:waterlevel} is used to ensure that with the resulted transmission schedule the boundary points, e.g., $a$ and $b$, are the desired BDP/BFPs. For the optimal BDP/BFPs set, the resulted transmission schedule must be feasible and the resulted water levels must satisfy the optimality conditions in Theorem \ref{thm:wf}.


We will consider the ``manually" generated set of BFP/BDPs which is called general BDP/BFPs set, where the BDP/BFPs are generated by constraining the battery be empty or fully-charged in some specific slots.
In contrast to the optimal BDP/BFPs set, the energy transmission obtained by \eqref{eq:swfc} based on a general BDP/BFPs set may not be feasible and/or optimal. Specifically, if the obtained transmission energy satisfies the constraints in \eqref{eq:nspcst}, we call the transmission schedule {\em feasible}; if the obtained transmission energy only violates the battery capacity constraints in some time slots, i.e, $B^k \leq B^{\max}$ for some $k$, we call the transmission schedule {\em semi-feasible}; otherwise, we call the transmission schedule {\em infeasible}. Note that both semi-feasible and infeasible transmission schedules are not feasible to the problem in \eqref{eq:nsp}-\eqref{eq:nspcst}.

For a specific case that, given a general BDP/BFPs set, by \eqref{eq:swfc}, the transmission schedule is feasible and the corresponding water levels satisfy the optimality conditions in Theorem \ref{thm:wf}, this general BDP/BFPs set can be considered as the optimal BDP/BFPs set and the corresponding transmission schedule is the optimal solution to the problem in \eqref{eq:nsp}-\eqref{eq:nspcst}. Therefore, in the following subsections, we want to compose such optimal BDP/BFPs set.

\subsection{Composing Optimal BDP/BFP Set}
Denote a general BDP/BFPs set as ${\cal X}=\{(k, \textrm{type of } k)\;|\; k\textrm{ is a BDP or BFP}\}$ where each element in $\cal X$ is sorted by ascending order of $k$, e.g., a set that contains only the default BDPs is $\{(0,\textrm{BDP}),(K,\textrm{BDP})\}$. To obtain ${\cal X}^*$, starting from ${\cal X}=\{(0,\textrm{BDP})\}$, we can iteratively append the next optimal BDP or BFP to ${\cal X}$ until $(K,\textrm{BDP})$ is added, i.e., we consider the generated BDP/BFP set as ${\cal X}^*$. To identify if a specific time slot should be a BDP or BFP in ${\cal X}^*$, we recursively perform the following two operations on a segment between $(a,\textrm{type of }a)$ and $(b,\textrm{type of }b)$: {\em Forward Search} and {\em Backward Search}.

For the forward search operation, we find the largest $(k,\textrm{type of } k) \in \{(a+1,\textrm{BDP}),(a+2,\textrm{BDP}),\ldots,(b-1,\textrm{BDP}),(b,\textrm{type of }b)\}$ such that the transmission schedule of the segment $[a+1,k]$, which is calculated by \eqref{eq:swfc}, is feasible or semi-feasible. If it is feasible, add $(k,\textrm{BDP})$ to $\cal X$ and continue the forward search for the segment between $(k,\textrm{BDP})$ and $(K,\textrm{BDP})$; if it is semi-feasible, we perform backward search for the segment between $(a,\textrm{type of }a)$ and $(k,\textrm{BDP})$.

For the backward search operation, we find and eliminate the largest $B^{\max}$-violation point, i.e., the largest $k\in[a+1,b]$ such that $B^k>B^{\max}$. Specifically, we first obtain the transmission schedule of the segment $[a+1,b]$ using \eqref{eq:swfc} and find out the largest $B^{\max}$-violation point $k$. Then, setting $k$ as a BFP, we obtain the transmission schedule of the segment $[a+1,k]$ using \eqref{eq:swfc} again. If the transmission schedule is infeasible, we perform forward search on the segment between $(a, \textrm{type of } a)$ and $(k,\textrm{BFP})$; if the transmission schedule is feasible, add $(k,\textrm{BFP})$ to $\cal X$ and perform forward search on the segment between $(k,\textrm{BFP})$ and $(K,\textrm{BDP})$; otherwise, we perform backward search on the  segment between $(a, \textrm{type of } a)$ and $(k,\textrm{BFP})$.

The steps involved in these two operations are described below. We set the initial BDP/BFPs set as ${\cal X}=\{(0, \textrm{BDP})\}$ and apply the forward search operation on the segment between $(0, \textrm{BDP})$ and $(K, \textrm{BDP})$ to get the optimal set of BDPs and BFPs.\\
\begin{minipage}[h]{6.5 in}
\rule{\linewidth}{0.3mm}\vspace{-.05 in}
{\bf {\footnotesize Algorithm 2 - Algorithm for finding optimal BDPs and BFPs}}\vspace{-.1in}\\
\rule{\linewidth}{0.2mm}
{ {\small
\begin{tabular}{ll}
	1:& Algorithm: Run {\sf Forward Search} on $\big((0,\textrm{BDP}),(K,\textrm{BDP})\big)$\\
    2:&Subroutine 1 - {\sf Forward Search $\big((a,\textrm{type of }a),(b,\textrm{type of }b)\big)$}\\
    & If $a=K$, the search is complete. \\
    &  For $(k_1, \text{ type of } k_1) \in \{ (a+1,\textrm{BDP}),\ldots,(b-1,\textrm{BDP})$, $(b,\textrm{type of }b)\}$\\
    & let $k$ = the largest $k_1\in (a,b]$ such that the transmission  schedule from $a+1$ to $k_1$ calculated by \eqref{eq:swfc} is  \\
    &  feasible or semi-feasible\\
    & \quad - if feasible, add $(k,\textrm{BDP})$ to ${\cal X}$ \\
    &\quad \quad and {\sf Forward Search $\big((k,\textrm{type of } k),(K,\textrm{BDP})\big)$}\\
    & \quad - if semi-feasible, {\sf Backward Search $\big((a,\textrm{type of } a),(k,\textrm{BDP})\big)$}\\
    3:& Subroutine 2 - {\sf Backward Search $\big((a,\textrm{type of }a),(b,\textrm{type of }b)\big)$}\\
    & Let $k$ = the largest $B^{\max}$-violation point in $(a,b]$\\
    & For the transmission schedule calculated by \eqref{eq:swfc} for segment $[a+1,k]$ where $k$ is BFP\\
    & \quad - if feasible, add $(k,\textrm{BFP})$ to ${\cal X}$ \\
    &\quad \quad and {\sf Forward Search $\big((k,\textrm{BFP}),(K,\textrm{BDP})\big)$}\\
    & \quad - if semi-feasible, {\sf Backward Search $\big((a,\textrm{type of }a),(k,\textrm{BFP})\big)$}\\
    & \quad - if infeasible, {\sf Forward Search $\big((a,\textrm{type of }a),(k,\textrm{BFP})\big)$}\\
\end{tabular}}}\\
\rule{\linewidth}{0.3mm}
\end{minipage}\vspace{.005 in}\\

Note that, Algorithm 2 is a recursive algorithm, in which a BDP is added by the forward search while a BFP is added at the end of a consecutive recursion of the backward search. Specifically, readdressing the definition of the water level of a segment in \eqref{eq:waterlevel}, when a BDP is added, it is ensured that the water level of the segment between the last BDP/BFP and the newly added BDP be lower than that of the segment between the newly added BDP and the next BDP/BFP that will be added in the subsequent recursion. Moreover, when a BFP is added, the opposite is ensured.

Note that Algorithm 2 is implemented by recursively performing the forward search and the backward search, based on the water-filling operation. In the recursive process, the starting point $a$ can only increase from $0$ to $K$ while the ending point $b$ can only decrease from $K$ to $a$ for each starting point $a$. Since the complexity of the traditional water-filling algorithm is ${\cal O}(1)$, the complexity of Algorithm 2 can be further bounded by ${\cal O}(K^2)$.

\subsection{Proof of Optimality}
In this subsection, we show that the BDP/BFP set composed by Algorithm 2 is the optimal BDP/BFP set, corresponding to the optimal transmission schedule for the problem in \eqref{eq:nsp}-\eqref{eq:nspcst}.

First we show that every BDP point picked by Algorithm 2 is in the optimal BDP/BFP set.

\begin{proposition}\label{pp:bdp}
Suppose we use \eqref{eq:swfc} to calculate the transmission schedule, and let $a,c\ (a<c)$ be BDP/BFPs such that the transmission schedule of the segment $[a+1,c]$ is infeasible. If $b\in[a+1,c]$ is the largest BDP and the transmission schedules of segments $[a+1,b]$ and $[b+1,c]$ are both feasible, then $b$ must be in the optimal BDP/BFP set for the segment $[a+1,c]$.
\end{proposition}
\begin{IEEEproof}
Since the transmission schedules for $[a+1,b]$ and $[b+1,c]$ are both feasible, the battery capacity constraints do not take effect. This result then follows from the staircase water-filling algorithm and the optimality proof [Lemma 2]  in \cite{OEAWCE} since the battery capacity constraint and the maximum energy consumption constraint are not dominant.
\end{IEEEproof}

Extending the above proposition to the case with finite battery capacity, we get the following corollary, showing the optimality of BDP added by Algorithm 2.

\begin{corollary}\label{cl:bdp}
Suppose that ${\cal X}^*$ is the optimal BDP/BFP set obtained by Algorithm 2, and let $a,b,c$ be three adjacent points in ${\cal X}^*$ where $b$ is a BDP added by the forward search operation. Then the water level of the segment $[a+1,b]$ is lower than that of the segment $[b+1,c]$.
\end{corollary}
\begin{IEEEproof}
Following Algorithm 2, for any segment between two adjacent points added to ${\cal X}^*$, there must exist a feasible transmission schedule calculated by \eqref{eq:swfc}. By Proposition \ref{pp:bdp}, $b$, which is added by the forward search, is the optimal BDP for the segment $[a+1,c]$. Then by Theorem \ref{thm:wf}, the single water level of the segment $[a+1,b]$ is lower than that of the segment $[b+1,c]$.
\end{IEEEproof}

\begin{figure}[!hbp]
\centering
\includegraphics[width=.9\textwidth]{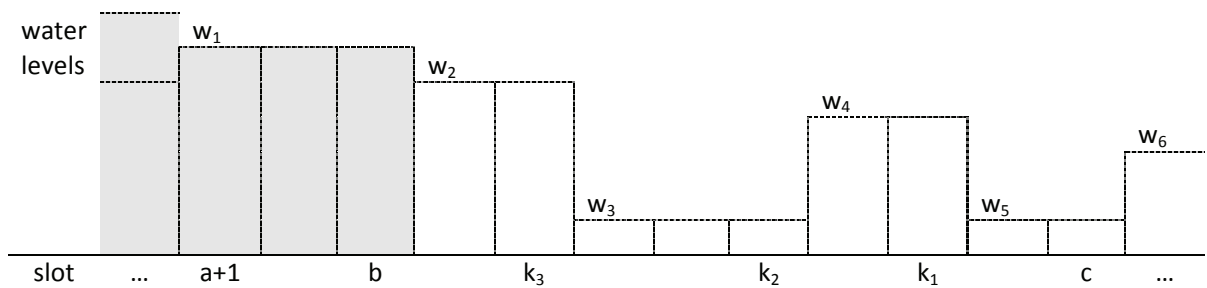}
\caption{An example of a consecutive recursion backward search. At the top level, the backward search is performed on $[a+1,c]$, where $a$ is BDP/BFP and $c$ is BDP, and $k_1$ is the largest $B^{\max}$-violation points (``semi-feasible" case). At the second, third and fourth levels, the backward search is performed on $[a+1,k_1]$, $[a+1,k_2]$, $[a+1,k_3]$, and $[a+1,b]$, where $k_2$, $k_1$, and $b$ are the largest $B^{\max}$-violation points in the second, third, and fourth levels (``semi-feasible" case) and then set as BFP in the next level, respectively. At the bottom (the fifth) level, the transmission schedule of $[a+1,b]$ is feasible (``feasible" case) and then $b$ is added as an optimal BFP. The water level $w_1$ is higher than any of the other water levels and the transmission schedule for the shaded area is feasible.}\label{fg:l3_1}
\end{figure}

Before we show the optimality of the BFP, we consider a consecutive recursion of backward search, where the top level is started by a forward search, each non-bottom level corresponds the ``infeasible" case, i.e., another backward search is called for the next recursion, and the bottom level is a ``feasible" case, i.e., a BFP is added to the BDP/BFPs set and then the forward search will be called for the next recursion. A simple example is shown in Fig. \ref{fg:l3_1} where $[a+1,c]$ is the segment of interest.

\begin{figure}[!hbp]
\centering
\includegraphics[width=.9\textwidth]{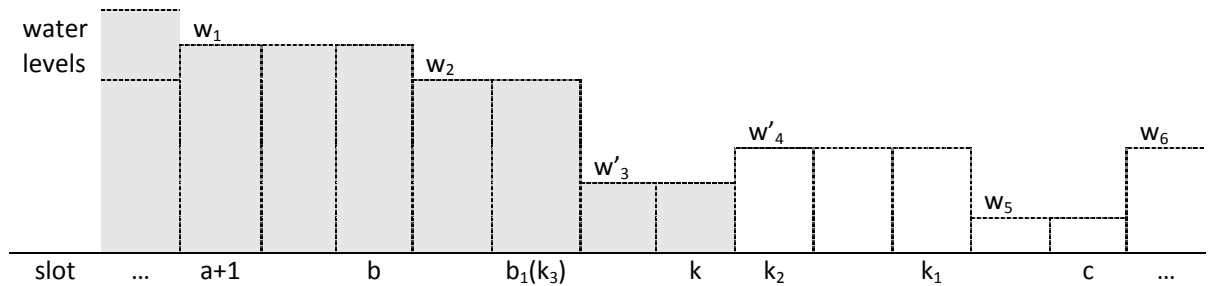}
\caption{An example of the existence of the BFPs and BDP proved by Lemma \ref{lm:bfp}, based on the example in Fig. \ref{fg:l3_1} (Assume the transmission schedule of $[b+1,b_1]$ is feasible.). We attempt to reschedule the energy for the segment $[b_1+1,k_1]$  ($[k_3+1,k_1]$ in Fig. \ref{fg:l3_1}) to form a constant water level under a feasible schedule; however, the ``best" result of the energy rescheduling is forming a new water level $w'_3$ for the segment $[b_1+1,k]$ under a feasible schedule, where $k$ is a BDP and $w_3<w'_3<w'_4<w_4$. (The detailed method is provided in the proof of Lemma \ref{lm:bfp}.) Also, for the shaded area, the transmission schedule is feasible. }\label{fg:l3_2}
\end{figure}

In this recursion process, the largest $B^{\max}$-violation point in each recursion level (e.g., $b,k_1,k_2,k_3$ in Fig. \ref{fg:l3_1}) is decreasing over the recursion process and iteratively set as BFP; the transmission schedule of the segment between the starting point and the added BFP (e.g., $b$ in Fig. \ref{fg:l3_1}), a.k.a., the largest $B^{\max}$-violation point in the level before the bottom level, is feasible. With the starting point of the consecutive backward search recursion process (e.g., $a$ in Fig. \ref{fg:l3_1}), and the added BFP (e.g., $b$ in Fig. \ref{fg:l3_1}), the following lemma proves the existence of another series of BFPs and a BDP. Specifically, for the existed another series of BFPs and a BDP, the water levels of the corresponding segment satisfy the optimal conditions of the BFPs in Theorem \ref{thm:wf}, showing in Fig. \ref{fg:l3_2} as a simple example.


\begin{lemma}\label{lm:bfp}
Suppose that, for a segment $[a+1,c]$ where $c$ is a BDP, the transmission schedule calculated by \eqref{eq:swfc} is semi-feasible.  Further suppose that $b$ is the BFP, which is added at the bottom-level of the consecutively recursive backward search process over the segment $[a+1,c]$. Then, we can always find a series of BFPs $b_1<b_2<\ldots<b_l\ (l\geq0,b_1>b)$ and a BDP $k\ (c\geq k>b_l)$ such that, the single water levels for $[a,b],[b+1,b_1],[b_1+1,b_2],\ldots,[b_l+1,k]$ are decreasing over $b_i$ and the corresponding transmission schedules are feasible.
\end{lemma}
\begin{IEEEproof}
The proof is provided in Appendix B.
\end{IEEEproof}

Lemma \ref{lm:bfp} considers the case that an optimal BFP is added to the BDP/BFPs set by Algorithm 2, denoted as $b$. It also shows that there exists a series of BFPs $b_i$ and a BDP $k$ such that the water levels of the segment separated by $b$ and $b_i$ decrease over $b_i$. Denoting the starting point for the consecutive backward search recursion process as $a$, we know that $b$ and $b_i$ are the optimal BFPs for the segment $[a+1,k]$ by Theorem \ref{thm:wf}. The following proposition shows the optimality of the BFP.

\begin{proposition}\label{pp:bfp}
Suppose that ${\cal X}^*$ is the optimal BDP/BFP set obtained by Algorithm 2. Let $a,b,c$ be three adjacent points in ${\cal X}^*$, where $b$ is the BFP added by the backward search operation. Then, the single water level of segment $[a+1,b]$ is higher than that of segment $[b+1,c]$.
\end{proposition}
\begin{IEEEproof}
We first focus on the subsegments between $b_i$ and $k$, whose existences have been proved in Lemma \ref{lm:bfp}. By Theorem \ref{thm:wf}, we know that the BFP $b$ added by Algorithm 2 must be in the optimal BDP/BFP set for $[a+1,k]$.

We consider a longer segment $[a+1,K]$. No matter how we set the BDP/BFP after the slot $k$, the water level before $k$ cannot be increased since $k$ is a BDP - any increase of the water level before $k$ causes the corresponding transmission schedule to be infeasible. If the water level before $k$ decreases by setting BDP/BFP after $k$, the existing BFPs before $k$ must still be BFPs since no more energy can flow over the BFP point. Therefore, $b$ must also be in the optimal BDP/BFP set for $[a+1,K]$.

Since Algorithm 2 iteratively performs the forward search and backward search operations, if the next added BDP/BFP $c$ is a BFP, it should also be in the optimal BDP/BFP set for $[a+1,K]$; while if the next added BDP/BFP $c$ is a BDP, it should be larger than $k$.

Therefore, by Theorem \ref{thm:wf}, the single water level of segment $[a+1,b]$ must be higher than that of segment $[b+1,c]$.
\end{IEEEproof}

To summarize, Corollary \ref{cl:bdp} shows that for each BDP added by the forward search operation, the water level of the segment before the BDP is lower than that after the BDP; while Proposition \ref{pp:bfp} shows that, for each BFP added by the backward search operation, the water level of the segment before the BFP is higher than that after the BFP. By Theorem \ref{thm:wf}, we know that such BDPs and BFPs added by Algorithm 2 can satisfy the optimality conditions for the problem in \eqref{eq:nsp}-\eqref{eq:nspcst}, i.e., the water levels may only increase at BDPs and only decrease at BFPs. Thus, we arrive at the optimality of Algorithm 2.

\begin{theorem}
By performing Algorithm 2, the resulted BDP/BFPs set is the optimal set of BDPs and BFPs for the problem in \eqref{eq:nsp}-\eqref{eq:nspcst}, i.e., we can get the optimal transmission schedule by using \eqref{eq:swfc} to water-fill each segment between two adjacent points in the optimal BDP/BFP set with a constant water level.
\end{theorem}


\section{Iterative Dynamic Water-filing for MAC}
In Section III, we proposed a dynamic water-filling algorithm to efficiently solve the single-user problem given by \eqref{eq:nsp}-\eqref{eq:nspcst}. In this section, we extend this algorithm to solve the general multi-user problem in \eqref{eq:problem}-\eqref{eq:cst}.

For a multiple-access channel, we define the transmission schedule as ${\cal P}\triangleq\{\boldsymbol{p}_i\;|\;i\in{\cal N}\}$ and the wastage schedule as  ${\cal D}^*\triangleq\{\boldsymbol{D}^*_i\;|\;i\in{\cal N}\}$, where $\boldsymbol{D}^*_i$ is obtained by Algorithm 1. Denote
\begin{equation}
M({\cal D})\triangleq \max_{{\cal P}}C({\cal P})\ ,
\end{equation}
subject to:
\begin{equation*}
\left\{\begin{array}{l}
 E_n^k -  B_n^{\max} - \sum_{t=1}^{k}D_n^{t} \leq \sum_{t=1}^{k}p_n^{t} \leq E_n^k - \sum_{t=1}^{k}D_n^{t}  \\
0 \leq p_n^k \leq P_n\\
k \in {\cal K}, \ n\in{\cal N}
\end{array}
\right. .
\end{equation*}

We note that for the multi-user problem, {each user has an independent battery evolution process with their individual energy constraints,  and thus the feasible domain of transmission schedule for any transmitter does not depend on the other transmitters (the optimal value is a result of a joint optimization problem, but the domains of the transmission schedule are independent). Therefore, the analyses in the proof of Lemma \ref{lm:1}, Lemma \ref{lm:2}, and Theorem \ref{thm:oo} are not affected by the additional term in the logarithm.} Thus, we can extend Theorem \ref{thm:oo} to the MAC case, i.e.,
\begin{equation}
M({\cal D}^*) = \max_{{\cal P},{\cal D}}C({\cal P})
\end{equation}
subject to the constraints in \eqref{eq:cst} for all $k\in{\cal K}, n\in{\cal N}$.

Then, applying the optimal wastage ${\cal D}^*$, $M({\cal D}^*)$ is given as follows.
\begin{equation}\label{eq:macproblem}
\max_{{\cal P}}C({\cal P})
\end{equation}
subject to
\begin{equation}\label{eq:maccst}
\left\{\begin{array}{l}
0 \leq {\tilde E}_n^k - \sum_{t=1}^{k}p_n^{t} \leq B_n^{\max} \\
0 \leq p_n^k \leq P_n\\
\end{array}
\right.
\end{equation}
for all $k \in {\cal K}$ and $n\in{\cal N}$, where $ {\tilde E}_n^k \triangleq  E_n^k -  \sum_{t=1}^{k}{D^*}_n^{t}$. We will solve \eqref{eq:macproblem}-\eqref{eq:maccst} in this section.

Similar to the original problem in \eqref{eq:problem}-\eqref{eq:cst}, the problem in \eqref{eq:macproblem}-\eqref{eq:maccst} is a convex optimization problem with ${\cal O}(NK)$ constraints and solving it with the general convex tools still encounter the computational complexity issue. To efficiently obtain the optimal transmission schedule, we will develop an iterative dynamic water-filling algorithm based on Algorithm 2.

\subsection{Joint Optimal Single-User Solution}

Denoting $\bar {\cal P}_n\triangleq {\cal P}\backslash\boldsymbol{p}_n$ as the transmission schedules other than that of transmitter $n$,  from the perspective of transmitter $n$, the objective function in \eqref{eq:obj} can be rewritten as:
\begin{equation}
\tilde C_n(\boldsymbol{p}_n,\bar {\cal P}_n)  \triangleq \sum_{k\in{\cal K}}\log(1+p_n^kH_n^k+\sum_{i\in{\cal N}/n}p_i^kH_i^k)
\end{equation}
From the perspective of transmitter $n$, we can further form a single user problem:
\begin{equation}\label{eq:osp}
\max_{\boldsymbol{p}_n} \tilde C_n(\boldsymbol{p}_n,\bar {\cal P}_n)
\end{equation}
subject to the constraints in \eqref{eq:maccst} for all $k\in{\cal K}$ given $\bar {\cal P}_n$.

Then, we want to prove that the problem to obtain the MAC optimal transmission schedule in \eqref{eq:macproblem}-\eqref{eq:maccst} can be simplified to obtaining a group of jointly optimal single-user transmission schedule problem in \eqref{eq:osp}.

\begin{theorem}\label{thm:dc}
Given ${\cal D}^*$, the schedule given by ${\cal P}$  is the optimal solution to \eqref{eq:problem}-\eqref{eq:cst}, if and only if, for every transmitter $n\in{\cal N}$, $\boldsymbol{p}_n\in{\cal P}$ is the optimal solution to the problem in \eqref{eq:osp} given $\bar{\cal P}_n$.
\end{theorem}

The conclusion follows from the strict concavity of the objective function over single variables and joint concavity over all variables. We provide a simple proof for this case in Appendix C.

Moreover, the objective function in \eqref{eq:osp} can be further rewritten as
\begin{align}
&C_n(\boldsymbol{p}_n,\bar {\cal P}_n)\nonumber\\
&=\sum_{k\in{\cal K}}\log(1+\frac{p_n^kH_n^k}{1+\sum_{i\in{\cal N}/n}p_i^kH_i^k})\nonumber\\
&\quad\quad+\sum_{k\in{\cal K}} \log(1+\sum_{i\in{\cal N}/n}p_i^kH_i^k)\nonumber\\
& = \sum_{k\in{\cal K}}\log(1+p_n^k\tilde{H}_n^k(\bar{\cal P}_n)) + \sum_{k\in{\cal K}}\log(H_n^k/\tilde{H}_n^k(\bar{\cal P}_n)) \label{eq:objeq}\ .
\end{align}
where
\begin{equation}\label{eq:heq}
\tilde{H}_n^k(\bar {\cal P}_n)\triangleq{H_n^k}/({1+\sum_{i\in{\cal N}/n}p_i^kH_i^k})\ .
\end{equation}
Note that, if $\bar{\cal P}_n$ is fixed, the second term in \eqref{eq:objeq} becomes a constant. Then, we can have the following problem such that its optimal solution is the same as that to \eqref{eq:osp}:
\begin{equation}\label{eq:esp}
\max_{\boldsymbol{p}_n}  \sum_{k\in{\cal K}}\log(1+p_n^k\tilde{H}_n^k(\bar{\cal P}_n))
\end{equation}
subject to the constraints in \eqref{eq:maccst} for all $k\in{\cal K}$ given $\bar {\cal P}_n$, where $\tilde{H}_n^k(\bar {\cal P}_n)$ is defined in \eqref{eq:heq}.

Specifically, the problem in \eqref{eq:esp} and the single-user problem discussed in section III have the same structure and therefore we may apply Algorithm 2 to solve the problem in \eqref{eq:esp}.

\subsection{Iterative Dynamic Water-Filling Algorithm}
To obtain the optimal schedule to the problem in \eqref{eq:problem}-\eqref{eq:cst}, we propose the following iterative algorithm and then show its optimality.\\
\begin{minipage}[h]{6.5 in}
\rule{\linewidth}{0.3mm}\vspace{-.05in}
{\bf {\footnotesize Algorithm 4 - Iterative dynamic water-filling algorithm}}\vspace{-.1in}\\
\rule{\linewidth}{0.2mm}
{ {\small
\begin{tabular}{ll}
	\;1:&  Initialization\\
	\;& $m=1$, ${\cal P}=\boldsymbol{0}$, $V^{(0)}=0$, specify the maximum iterations $M$\\
	\;2:& Local Wasting Schedule\\
	\;& Obtain $\boldsymbol{D}_n^*$ by the Algorithm 1 for all $n\in{\cal N}$\\
	\;3:& Local Energy Schedule\\
	\;&\textbf{\bf FOR} $n\in{\cal N}$\\
		\;&\quad Calculate $\tilde{H}^k(\bar {\cal P}_n)$ and solve \eqref{eq:esp} to update ${\boldsymbol{p}_n}$ in  ${\cal P}$ by Alg. 2\\
	\;&\textbf{\bf ENDFOR}\\
	\;4& Evaluation\\
	\;& $V^{(m)}=C({\cal P})$\\
	\;& \textbf{\bf IF} $|V^{(m)}-V^{(m-1)}|> 0$ and $m\leq M$\\
		\;&\quad \textbf{\bf GOTO STEP 3}\\
	\;& \textbf{\bf ENDIF}
\end{tabular}}}\\
\rule{\linewidth}{0.3mm}
\end{minipage}\vspace{.01 in}\\

The basic idea of this algorithm is to successively solve the optimization problem  in \eqref{eq:esp}  with  $\tilde{H}_n^k(\bar {\cal P}_n)$ to update $\cal P$ until $C({\cal P})$ converges. Specifically, in each iteration, with  $\tilde{H}_n^k(\bar {\cal P}_n)$, the algorithm successively solves  $S(\boldsymbol{D}^*_n)$ for $n \in {\cal N}$ to update the corresponding  $\boldsymbol{p}_n\in{\cal P}$. When the objective function $C({\cal P})$ converges, the solution given by $\cal P$ is the optimal solution to the problem in \eqref{eq:problem}-\eqref{eq:cst}. To show that the schedule obtained by Algorithm 4 is optimal, we need to verify that it satisfies the conditions in Theorem \ref{thm:dc}.

First, since the objective function in  \eqref{eq:esp} is strictly concave, then we have
\begin{proposition}\label{pp:unique}
Given $\bar{\cal P}_n$, the optimal solution to \eqref{eq:esp} with $\tilde{H}_n^k(\bar {\cal P}_n)$ is unique.
\end{proposition}

Note that, given ${\cal D}^*$, the alternating optimization  procedure in Algorithm 4 leads to a non-decreasing $C({\cal P})$. Since the objective value of \eqref{eq:problem} is bounded, a convergent point of $C({\cal P})$ can be achieved by Algorithm 4. The optimality of the convergent point is  shown in the following theorem.

\begin{theorem}\label{thm:opt}
{At the convergent point of Algorithm 4,  we obtain a transmission schedule ${\cal P}=[\boldsymbol{p}_1,\boldsymbol{p}_2,\ldots, \boldsymbol{p}_N]$ such that,
for each transmitter $n\in{\cal N}$, $\boldsymbol{p}_n$ is the optimal solution to the problem in \eqref{eq:osp} given $\bar{\cal P}_n={\cal P}\slash \boldsymbol{p}_n$.}
\end{theorem}
\begin{IEEEproof}
We denote ${\cal P}=\{\boldsymbol{p}_1,\boldsymbol{p}_2,\ldots,\boldsymbol{p}_N\}$ as the transmission schedule obtained by Algorithm 4 and denote $V=C({\cal P})$ as the convergent value, i.e., $V$ is the convergent point of the Algorithm 4 which is achieved by $\cal P$ and ${\cal D}^*$.

In Algorithm 4, \eqref{eq:esp}  is solved successively for $n = 1,\cdots N$ to update the corresponding $\boldsymbol{p}_n$ in $\cal P$ until $V$ converges. After $V$ converges, if we continue to update ${\cal P}$ by solving \eqref{eq:esp}, we can still obtain an optimal solution for transmitter $n$, denoted by $\boldsymbol{p}'_n$. Since $\cal P$ has already achieved convergence, we have $V=C( \boldsymbol{p}_n\cup\bar{\cal P}_n) = C(\boldsymbol{p}'_n\cup\bar{\cal P}_n)$, i.e., $\boldsymbol{p}_n$ is also the optimal solution to  \eqref{eq:osp}, as well as $\boldsymbol{p}'_n$.  By Proposition \ref{pp:unique}, the optimal solution to  \eqref{eq:osp} is unique and we must have $\boldsymbol{p}_n=\boldsymbol{p}'_n$, i.e., $\cal P$ also converges.

Moreover, $\cal P$ is generated by successively solving \eqref{eq:esp}. In particular, if $\cal P$ has converged, for any $\boldsymbol{p}_n$ and $\bar{\cal P}_n$ in ${\cal P}$, we have that  $\boldsymbol{p}_n$ achieves the optimality of  \eqref{eq:osp} given $\bar{\cal P}_n$ for all $n\in{\cal N}$. Then, by Theorem \ref{thm:dc}, we see that the obtained $\cal P$ is the optimal solution to the problem in \eqref{eq:problem}-\eqref{eq:cst}.
\end{IEEEproof}

In \cite{IWGVMC}, a result is provided on the gap between the converged solution and the solution after the first iteration for an iterative water-filling algorithm. To obtain a similar result for Algorithm 4, we reformulate the problem in \eqref{eq:problem}-\eqref{eq:cst} to the following form:
\begin{equation}
\left\{
\begin{array}{ll}
&\min_T -\log T\\
\textrm{s.t.}&T \leq  1+\sum_{i\in{\cal N}}p_i^kH_i^k\\
&0 \leq {\tilde E}_n^k - \sum_{t=1}^{k}p_n^{t} \leq B_n^{\max} \\
&0 \leq p_n^k \leq P_n\\
\end{array}
\right. \ ,
\end{equation}
for all $n\in{\cal N},k\in{\cal K}$. We then form its Lagrangian dual, and use the fact that the difference between the primal and dual objectives, the so-called duality gap, is a bound on the difference between the primal objective and the optimum. Thus, following similar steps as in \cite{IWGVMC}, we can characterize the gap between the dual objective function and the primal objective function at the end of the first full iteration, then obtain the following theorem.
\begin{theorem}
After the first iteration of Algorithm 4, the objective value is at most $(N-1)K/2$ nats away from the optimal value.
\end{theorem}

Note that, in each iteration of Algorithm 4, for each transmitter Algorithm 2 is called, whose complexity is ${\cal O}(K^2)$. Hence the complexity of Algorithm 4 is ${\cal O}(NK^2)$ per iteration.
\subsection{Suboptimality of TDMA}
It is shown in \cite{ICPCSM} that, without energy harvesting, TDMA is the optimal energy scheduling strategy for multiple-access channel with average power constraints. Next we show that in a finite-horizon energy harvesting system with finite battery capacity, the TDMA strategy is no longer optimal.

To characterize the optimal schedule, we need to focus on the KKT conditions of the problem given by \eqref{eq:problem}-\eqref{eq:cst} for each $n\in{\cal N}$. Note that the K.K.T. conditions of the MAC problem can be rewritten by adding each subscript $n\in{\cal N}$ to all dual variables in the corresponding single-user K.K.T. condition expect for \eqref{kkt:wf}, which is replaced by
\begin{align}
\gamma^k \triangleq 1+\sum_{i\in{\cal N}}p_i^kH_i^k
&= \frac{H_1^k}{v_1^k - u_1^k + \alpha_1^k - \beta_1^k}\nonumber\\
& =\frac{H_2^k}{v_2^k - u_2^k + \alpha_2^k - \beta_2^k}
\nonumber\\&= \ldots \  \ldots \nonumber\\&=\frac{H_N^k}{v_N^k - u_N^k + \alpha_N^k - \beta_N^k}\label{eq:TDMA}\ ,
\end{align}
for all $k\in{\cal K}$.

Since $H_n^k$ is a constant, the equality of \eqref{eq:TDMA} must be maintained by the dual variables $u_n^k,v_n^k,\alpha_n^k,\beta_n^k$. Suppose that, we know the optimal $u_n^k,v_n^k,\alpha_n^k,\beta_n^k,\gamma^k$. In time slot $k$, for transmitter $n$ such that $\frac{1}{v_n^k-u_n^k}>\frac{\gamma^k}{H_n^k}$, we must have $\alpha_n^k > 0$, resulting in the transmitter transmitting at the maximum energy consumption $P_n$, i.e., $p_n^k=P_n$ by \eqref{kkt:c}, thus causing $\beta_n^k=0$ by \eqref{kkt:d}. However, if for any $n$, $\frac{1}{v_n^k-u_n^k}<\frac{\gamma^k}{H_n^k}$, we must have $\beta_n^k>0$, resulting in the transmitter sending at zero power, i.e., $p_n^k=0$ by \eqref{kkt:d}, thus causing $\alpha_n^k = 0$ by \eqref{kkt:c}.

Moreover, in time slot $k$, for any transmitter $n$ such that $\frac{1}{v_n^k-u_n^k}=\frac{\gamma^k}{H_n^k}$, we must have $\alpha_n^k=\beta_n^k$. If $\alpha_n^k=\beta_n^k=0$, by \eqref{kkt:c} and \eqref{kkt:d}, transmitter $n$ may access the channel at any possible energy consumption. If $\alpha_n^k=\beta_n^k>0$, there exists contradiction and there is no optimal solution in this case.

Through the above analysis, we know that, in each slot $k$,  for the transmitters whose water level (the height of this level is essentially determined by the harvested energy) is at least $\gamma^k/H_n^k$, the corresponding optimal transmission schedule allows those transmitters to transmit simultaneously. For example, in a slot, if the harvested energy is larger than the transmitters' battery capacity,  to avoid/reduce the battery overflow, these transmitters should transmit to achieve the optimality. However, if we insist that only one user can access the channel in each slot, as in TDMA, the corresponding transmission schedule must be suboptimal for the finite-horizon energy harvesting system with finite battery capacity.

\section{Numerical Results}

Suppose that there are $N=5$ users in the system.  We set the scheduling period as $K=20$ slots. For each user $n\in{\cal N}$, we set the initial energy $B_n^0 = 0$. Assume that the harvested energy $E_n^k$ follows a nonnegative truncated Gaussian distribution with mean $m$ and variance $v$, and that the channel fading parameter $H_n^k\sim \exp(1)$,  which corresponds to the magnitude-squared standard complex Gaussian channel gain. Moreover, the parameter setup for the transmitter is guided by the energy-constrained system that operates highly related to the energy harvesting with the limited buffer battery, e.g., the EnHants in \cite{WPDFUE}, and the unit of the energy (e.g., $m$, $B_n^{\max}$) is ``$10^{-2}$J'', the unit of the per-slot energy consumption (e.g, $P_n$, $p_n^k$) is ``$10^{-2}$J/slot'', and the SNR is treated as the value of $p_n^kH_n^k$ at the receiver after taking account into the path loss, antenna gain, and etc.

For comparison, we consider three simple scheduling strategies, namely, the {\em greedy policy}, the {\em balanced policy}, the {\em modified staircase water-filling}~\cite{OEAWCE}. The greedy policy tries to consume the harvested energy as much as possible in each slot, as calculated by \eqref{eq:greedyp}.  On the other hand, the balanced policy tries to consume the fixed amount of energy in each slot if available, where the fixed value is calculated by $\sum_{k=1}^K E_n^k / K$. Moreover, the modified staircase water-filling obtains the feasible energy schedule by restricting the maximum energy consumption and wasting the overflowed energy of the energy schedule that is obtained by the staircase iterative water-filling algorithm proposed in \cite{OEAWCE} for the single-user case; for the multi-user case, we iterate the modified staircase water-filling similarly as in Algorithm 4 until convergence or the maximum number of iterations is reached. Specifically, for the original ``staircase water-filling algorithm'' as proposed in  \cite{OEAWCE}, from the last water-filling ending point, we need to iteratively try to identify the longest segment such that using the water-filling algorithm gives a feasible transmission schedule.

\begin{figure}[!hbp]
\centering
\includegraphics[width=.9\textwidth]{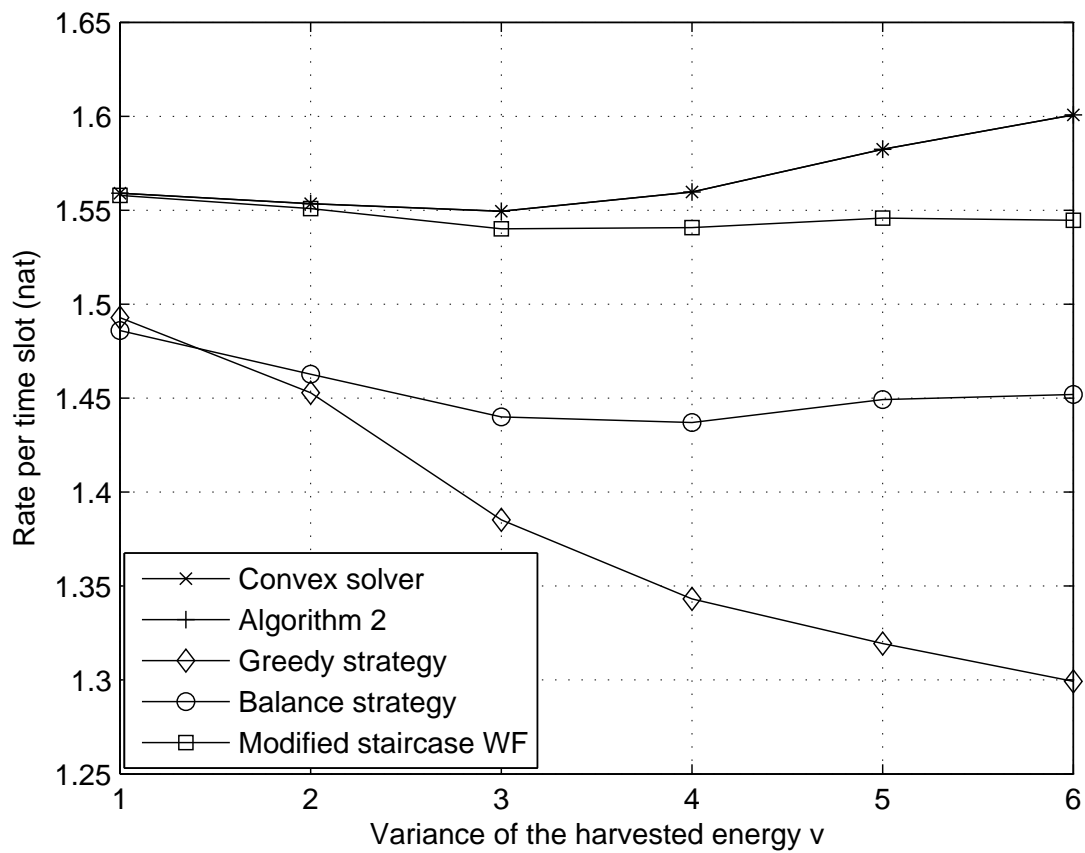}
\caption{Performance comparisons for the singe-user case in energy-constrained scenario.}\label{fg:p1}
\end{figure}

\begin{figure}[!hbp]
\centering
\includegraphics[width=.9\textwidth]{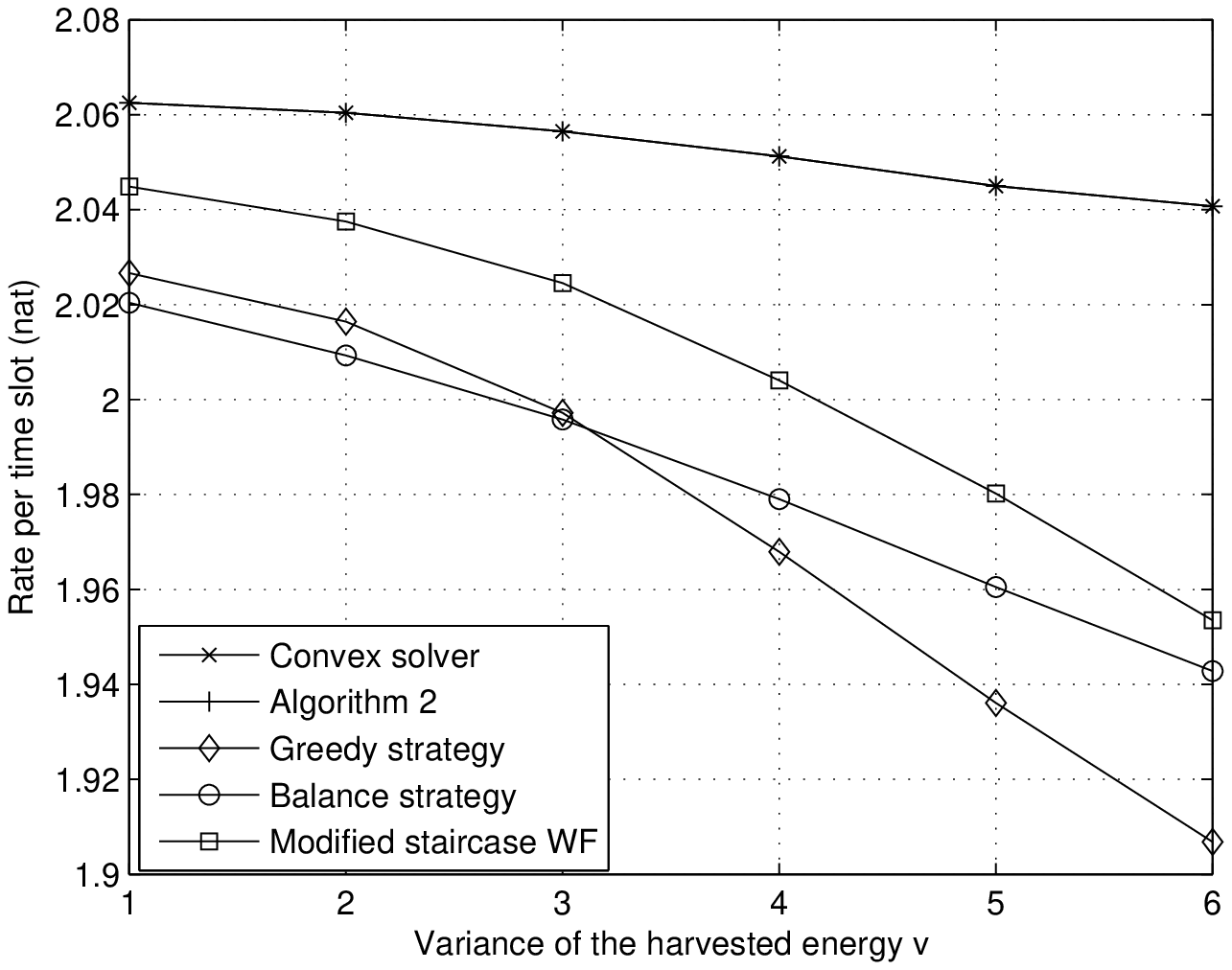}
\caption{Performance comparisons for the singe-user case in power-constrained scenario.}\label{fg:p2}
\end{figure}

\begin{figure}[!hbp]
\centering
\includegraphics[width=.9\textwidth]{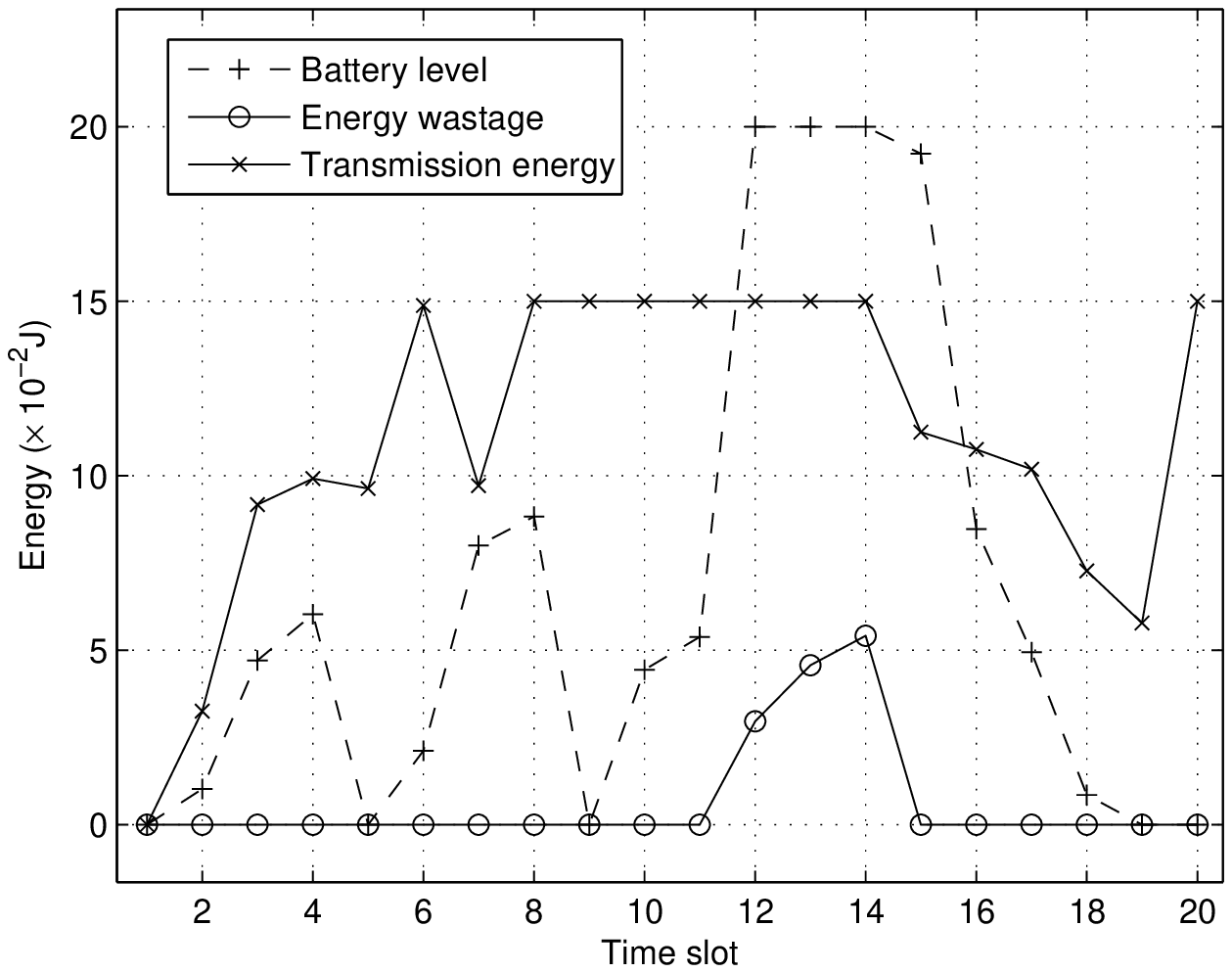}
\caption{An example of the energy schedule in power-constrained scenario.}\label{fg:example}
\end{figure}

We first consider two scenarios to evaluate the energy scheduling algorithms for the single-user case, namely, the  {\em energy-constrained} scenario, where $m = 5$, and the {\em power-constrained} scenario, where $m = 10$; moreover, in this simulation, all other parameters are set as $v=1,2,3,4,5,6$, $B_n^{\max}=20$, and $P = 15$. In the energy-constrained scenario, the harvested energy in each slot is usually below the maximum energy consumption constraint. Then, the transmission schedule is mainly constrained by the availability of the energy. However, in the power-constrained scenario, the harvested energy in each slot may reach the maximum energy consumption constraint and the transmission schedule is more constrained by the maximum energy consumption and battery capacity. We run the simulation $2000$ times and the average rates given by various scheduling strategies, as well as by the optimal schedule solved by the general convex solver, are shown in Fig.~\ref{fg:p1} and Fig.~\ref{fg:p2}. {Further, a numerical example of the battery level, the optimal energy wastage, and the optimal transmission energy is depicted in Fig.~\ref{fg:example} for power-constrained scenario.}



We see from Fig. \ref{fg:p1} and Fig. \ref{fg:p2} that for the single-user case the proposed algorithm and the optimal schedule solved by the general convex solver give the same performance as expected. Moreover, the modified staircase water-filling has worse performance than the optimal schedule, and the greedy policy and the balanced policy have worse performance than the modified staircase water-filling. In the energy-constrained scenario, most of the harvested energy can be buffered by the battery for later use. { Since the mean of the energy harvesting is small, with the well-buffered energy (most energy can be buffered instead of being wasted), increasing the variance of the harvested energy (distributed as a nonnegative truncated Gaussian random variable) leads to an increased total available energy and thus higher sum-rate, as shown in Fig. \ref{fg:p1}}. In the power-constrained scenario, {the mean of the harvested energy is high and the battery mostly operates near the fully-charged status.  As a result, increasing the variance of the harvested energy may only slightly increase the harvested energy but lead to larger energy fluctuation thus causing the battery to be fully-charged more frequently.} This causes more energy to be wasted, thus leading to lower available energy and sum-rate, as shown in Fig. \ref{fg:p2}. Moreover, since the modified staircase water-filling does not take into account  the maximum per-slot energy consumption and the maximum battery capacity, some harvested energy cannot be well utilized especially when the variance is large, resulting in the increased gap to the optimal schedule, as shown in Fig. \ref{fg:p1} and Fig. \ref{fg:p2}.

\begin{figure}[!hbp]
\centering
\includegraphics[width=.9\textwidth]{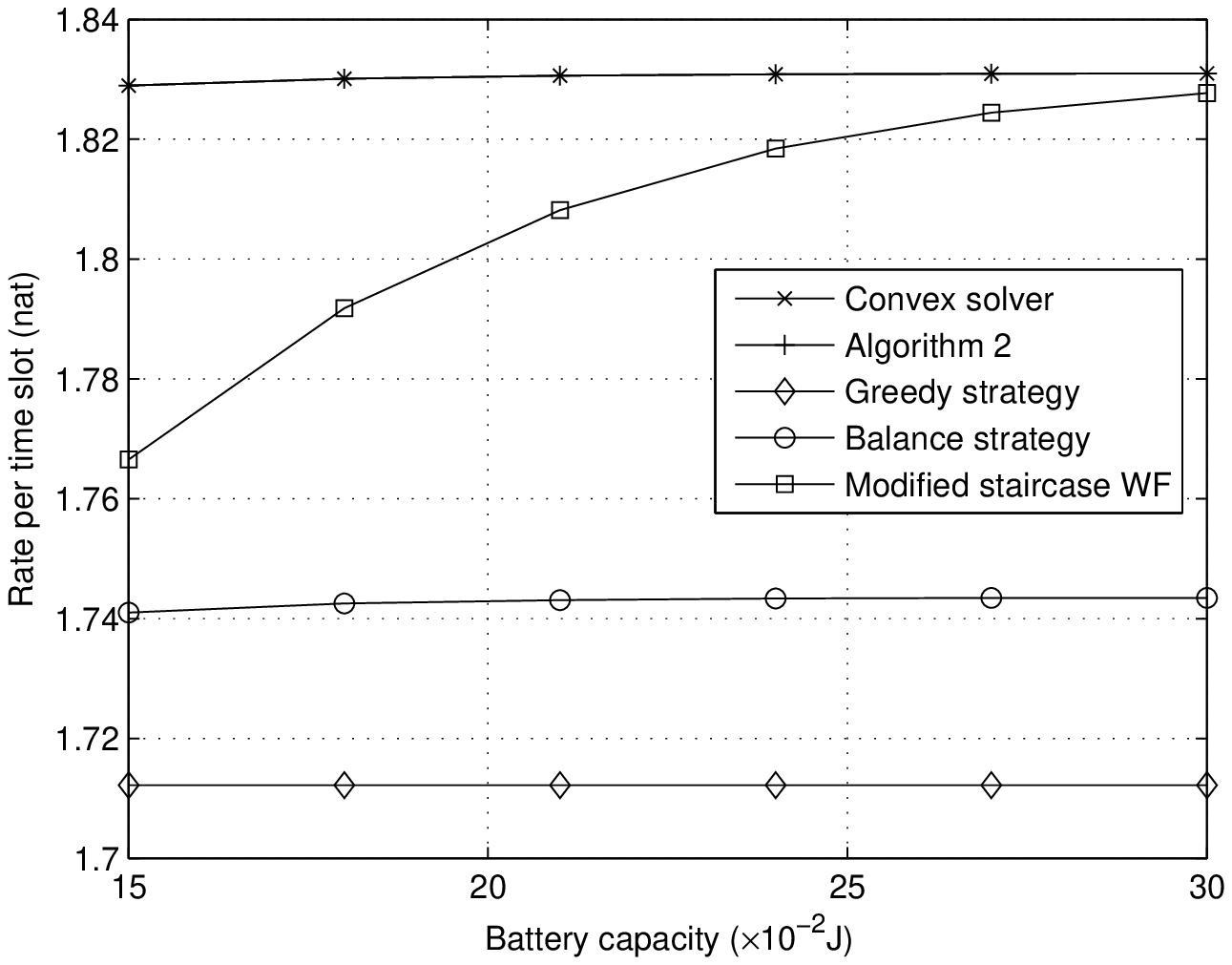}
\caption{Performance comparisons for the singe-user case in high-power scenario ($P=15$).}\label{fg:ap1}
\end{figure}

\begin{figure}[!hbp]
\centering
\includegraphics[width=.9\textwidth]{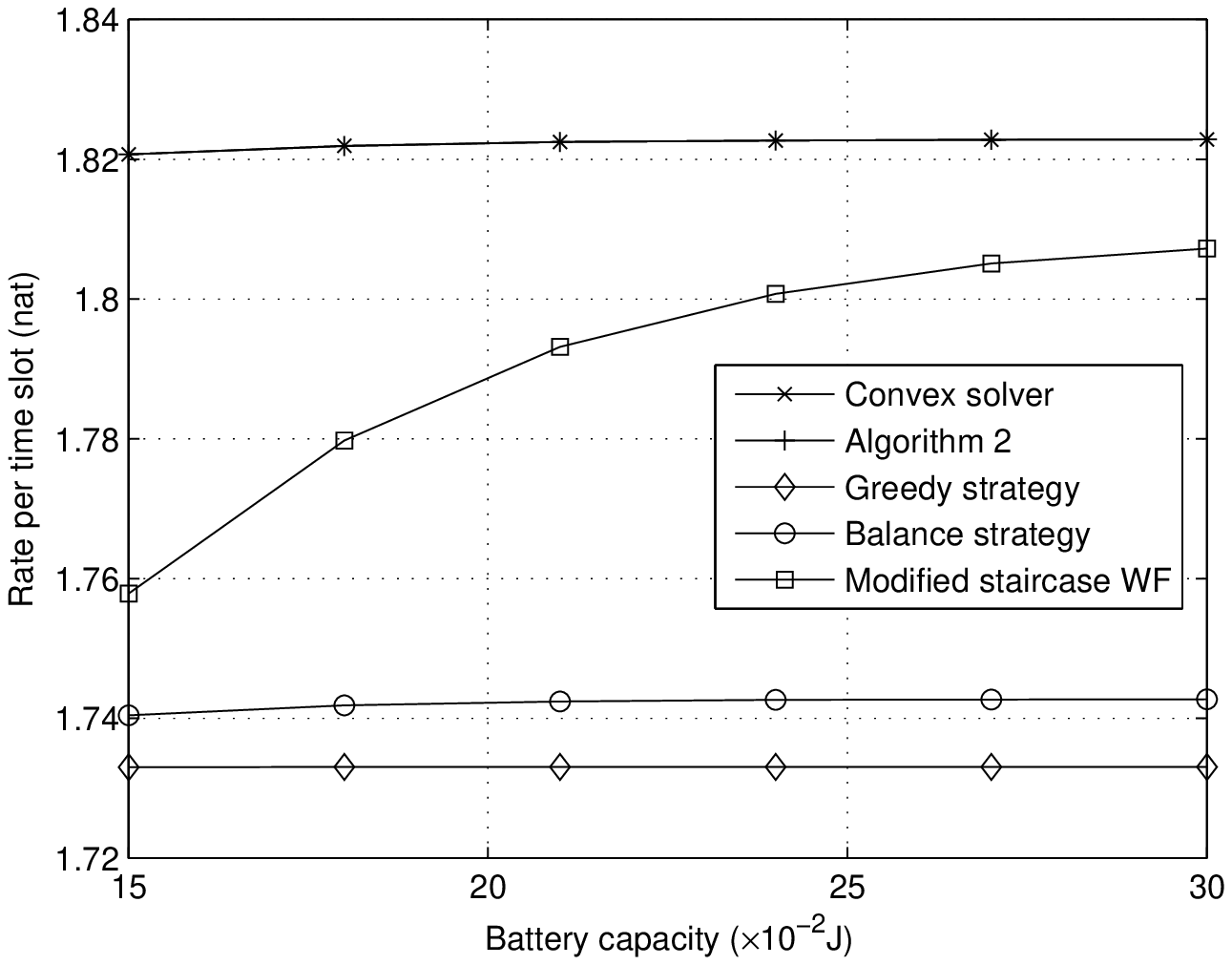}
\caption{Performance comparisons for the singe-user case in low-power scenario ($P=10$).}\label{fg:ap2}
\end{figure}

We next compare the performance of the various energy scheduling algorithms with two more scenarios  for the single-user case, namely, the  {\em high-power} scenario, where $P=15$, and the {\em low-power} scenario, where $P = 10$. For theses two scenarios, all other parameters are set as  $B^{\max} = 15,18,21,24,27,30$, $m=7.5$, and $v=3.5$.  As compared to the low-power scenario, the transmitter in high-power scenario has larger dynamic range to schedule the transmission energy and battery is more easily depleted. We run the simulation $2000$ times and the average rates given by various scheduling strategies, as well as by the optimal schedule solved by the general convex solver, are shown in Fig.~\ref{fg:ap1} and Fig.~\ref{fg:ap2}.



It is seen from Fig.~\ref{fg:ap1} and Fig.~\ref{fg:ap2} that  the proposed transmission schedule algorithm and the optimal schedule solved by the general convex solver achieve the same performance as expected and the other algorithms do not perform as well as the optimal one. In both scenarios, as the battery capacity increases, the performances of both algorithms improve, and the performance gap between the optimal schedule and the modified staircase water-filling becomes small. This is mainly because higher battery capacity may reduce the potential energy overflow and thus the energy schedule obtained by modified staircase water-filling algorithm approaches the optimal energy schedule.  Moreover, as expected, with the same battery capacity, the optimal performance in high-power scenario is better than that in low-power scenario, mainly caused by the different dynamic range for the transmission energy scheduling.

\begin{figure}[!hbp]
\centering
\includegraphics[width=.9\textwidth]{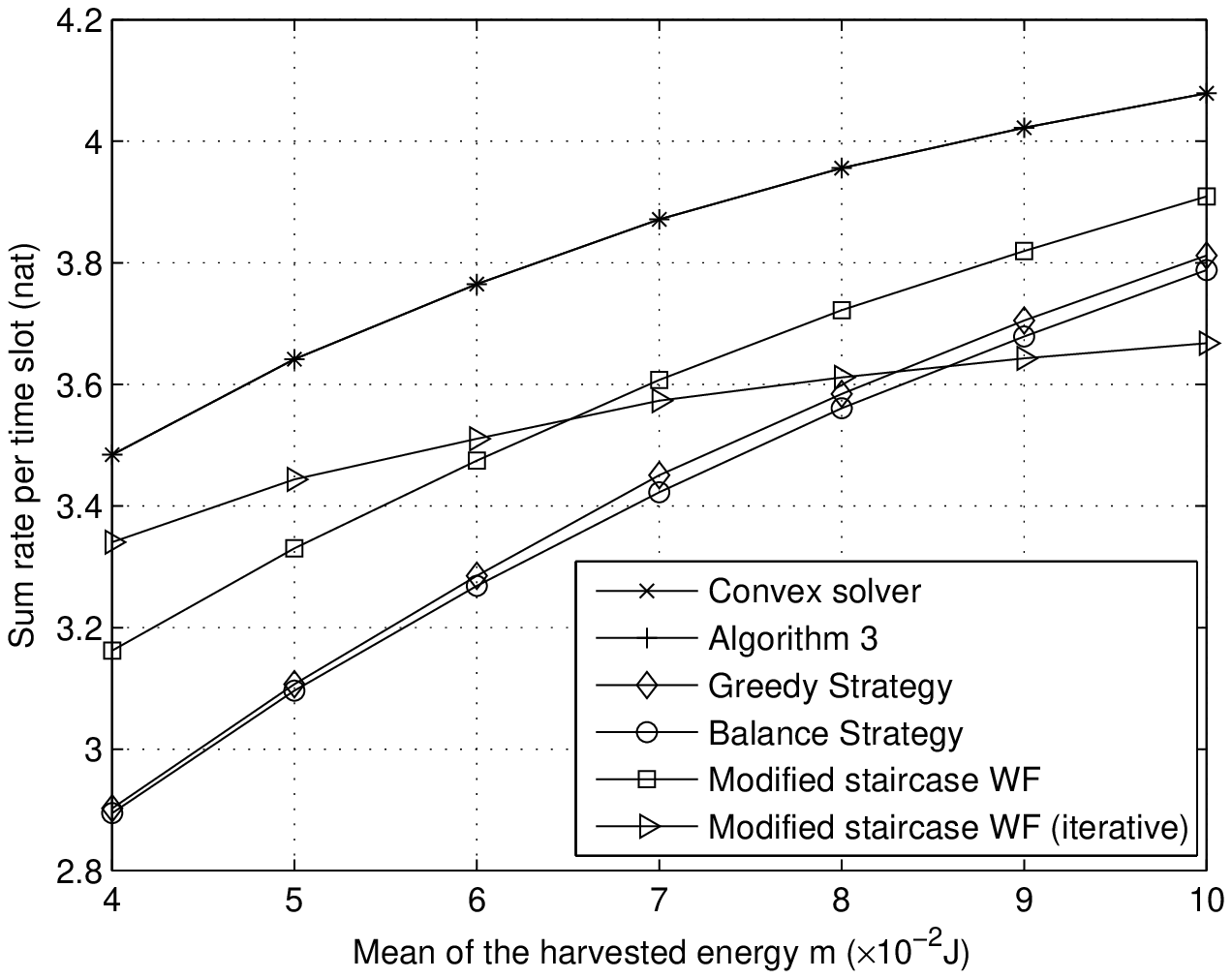}
\caption{Performance comparisons for the multi-user case ($v=3.5$).}\label{fg:p3}
\end{figure}

\begin{figure}[!hbp]
\centering
\includegraphics[width=.9\textwidth]{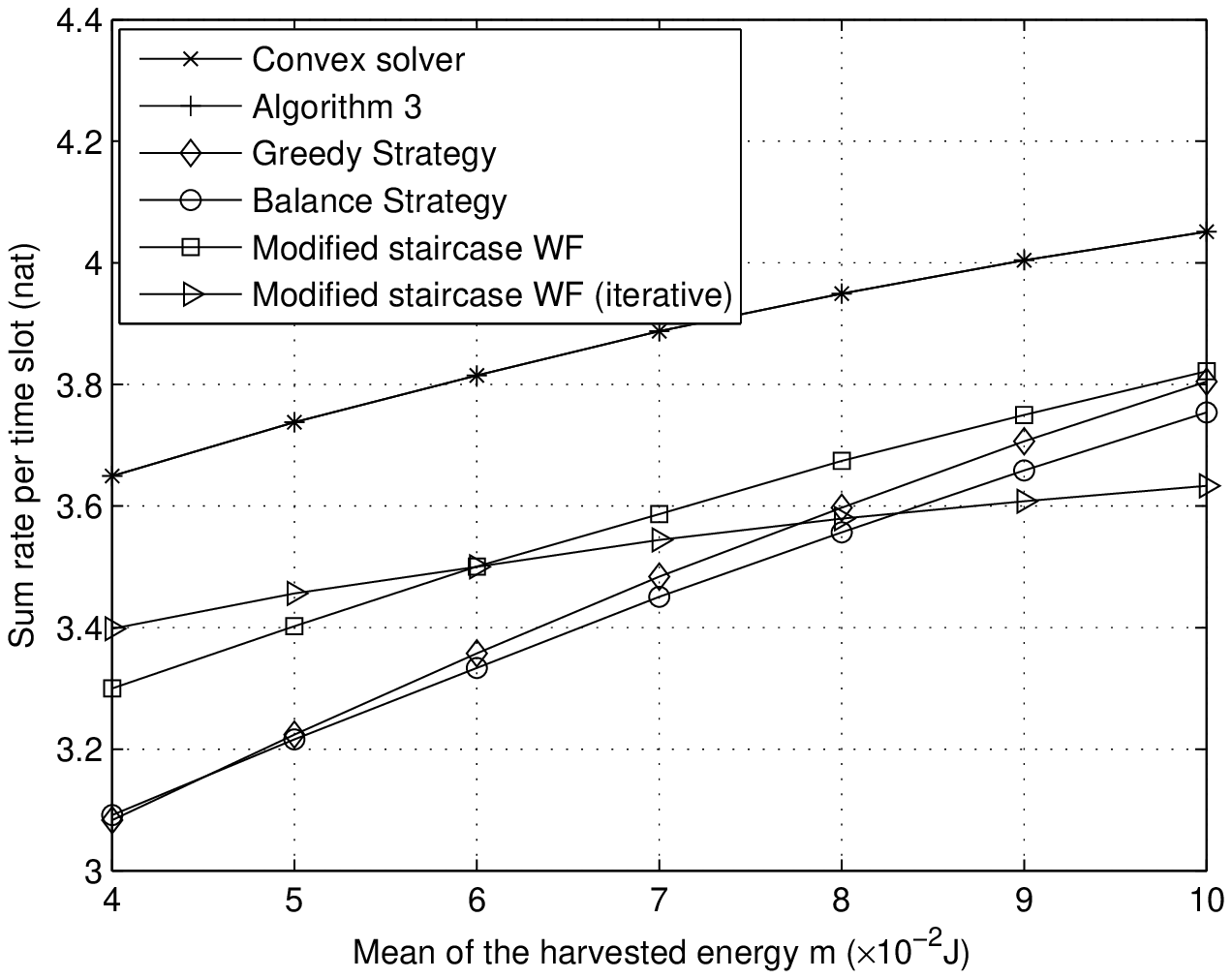}
\caption{Performance comparisons for the multi-user case ($v=8$).}\label{fg:ap3}
\end{figure}

\begin{figure}
\centering
\includegraphics[width=.9\textwidth]{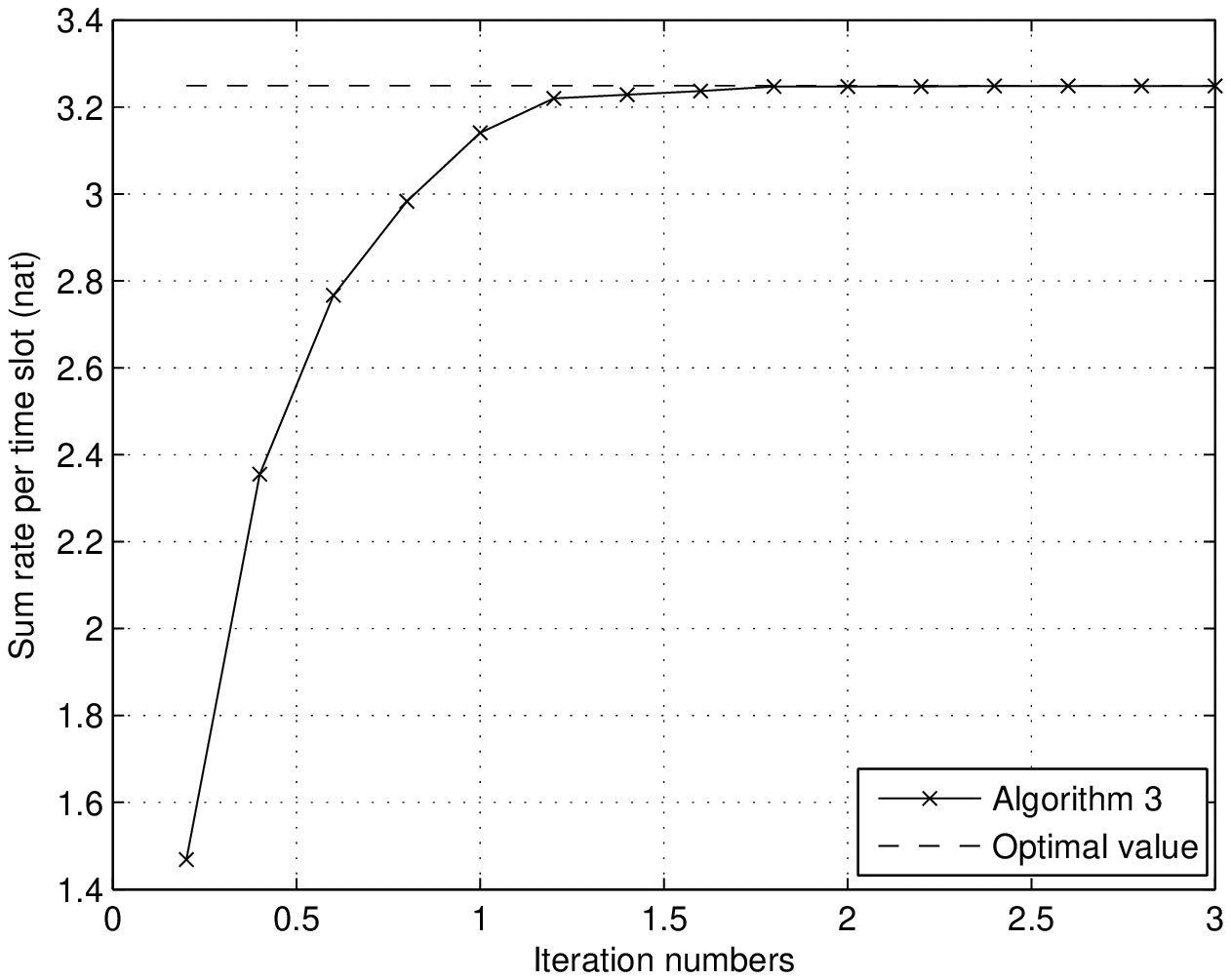}
\caption{The convergence of Algorithm 3 for the multi-user case ($m=5$, $v=8$).}\label{fg:p4}
\end{figure}

We then compare the performances of the various energy scheduling algorithms for the multi-user case for $v=3.5,8$, $m=5,6,7,8,9,10$, $B_n^{\max}=20$, and $P_n = 15$. In this scenario, the transmission schedule is mainly constrained by the energy availability when $m$ is small and constrained by the maximum energy consumption and battery capacity constraints when $m$ is large. We considered two versions of the modified staircase water-filling method in this simulation: the iterative version described earlier, and a non-iterative version where each transmitter independently obtains its energy schedule and the sum-rate is evaluated based on these independent energy schedules, similar to the greedy policy and the balanced policy. In Algorithm 4, we set the convergence threshold as $\epsilon = 10^{-5}$ and the maximum iteration number as $M = 50$. The average sum-rates given by various scheduling strategies are shown in Fig.~\ref{fg:p3} and Fig.~\ref{fg:ap3} for $v=3.8$ and $v=8$, respectively; the convergence behavior of Algorithm 3 is shown in Fig.~\ref{fg:p4}.

Fig.~\ref{fg:p3} and Fig.~\ref{fg:ap3} compare the performance of different strategies for the multi-user case with different $v$. Similar to the single-user case, it is seen that, for both $v=3.5$ and $v=8$ cases, the proposed energy scheduling algorithm and the optimal schedule solved by the general convex solver give the same performance, as expected. Also, the performance improves for all algorithms when $m$ increases. In addition, we observe that the performance of the iterative modified staircase water-filling is worse than other algorithms when $m$ is large. It is mainly because the iterative version tends to select the transmitter that has the best possible transmission energy to exclusively occupy the channel with large energy consumption in a slot. However, the maximum energy consumption constraint is not taken into account by the modified staircase water-filling, which leads to the truncation of the excessive portion of transmission energy, especially when $m$ is large.  Moreover, it is seen from Fig. \ref{fg:p4} that Algorithm 4 converges after only $2$ iterations.

\section{Conclusions}
In this paper, we have considered the energy scheduling problem for $N$-user fading multiple-access channel with energy harvesting over $K$ time slots. This problem is formulated as a convex optimization problem with ${\cal O}(NK)$ variables and  ${\cal O}(NK)$ constraints. To avoid the high computational complexity of the general convex solver, we have developed an efficient optimal solution, called the iterative dynamic water-filling algorithm. For the single-user case, this algorithm performs dynamic water-filling with different water levels between battery depletion and battery full-charged points. For the multi-user case, the algorithm iterates the dynamic water-filling algorithm for different users until convergence. Simulation results demonstrate that the proposed optimal energy scheduling algorithm converges in a few iterations and offers significant gain in term of sum-rate over various suboptimal schedules.

\appendices

\section{Proof of Lemma 2}
Following the energy harvesting and storing process, the harvested energy is stored in the battery and the available energy may be consumed and/or wasted in any slot - earlier or later - as long as the corresponding energy is still in the battery. So, for any feasible energy wastage schedule $\boldsymbol{D}=[D^1,D^2,\ldots,D^K]$ such that $\sum_{k\in{\cal K}}D^k = \sum_{k\in{\cal K}}{D^*}^k$, each can be generated from ${\boldsymbol{D}^*}$ by rescheduling part of the wasted energy ${D^*}^k$ to some previous slot for wasting as long as the corresponding energy is still stored in the battery, without violating the constraints in \eqref{eq:scst}. However, due to the causality of energy harvesting, the scheduled wasted energy can only be rescheduled to certain possible previous slots. From \eqref{eq:greedy}, we note that, in such possible rescheduled slots, the transmitter must transmit at maximum energy consumption; otherwise, we can increase the transmission energy to the maximum limit to reduce the energy wastage, contradicting the minimum energy wastage of ${\cal D}^*$. Therefore,  such reschedule operation does not change the feasible domain of transmission schedule $\boldsymbol{p}$ and thus the optimal solution with reschedule operation is the same as that without it. A typical example is illustrated in Fig.~\ref{fg:d1} and Fig.~\ref{fg:d2}.

\section{Proof of Lemma 3}
For a segment $[a+1,c]$ which is returned from the backward search, it has semi-feasible transmission schedule with a constant water level and there must exist a $B^{\max}$-violation point, e.g., slot $k_1$ in Fig.~\ref{fg:l3_1}. By enforcing the largest $B^{\max}$-violation point $k_1$ as a BFP, e.g., slot $k_1$ in Fig.~\ref{fg:l3_1}, we obtain the water levels for $[a,k_1]$ and $[k_1+1,c]$. As compared to the case when $K_1$ is $B^{\max}$-violation point, the battery level at the BFP point is reduced to $B^{\max}$, i.e., some of the energy is pulled back from the segment after the BFP to the segment before the BFP. From \eqref{eq:swfc}, we see that the water level of the segment before the BFP is increased while that of the segment after the BFP is decreased, as compared to the original water level of $[a+1,c]$. Since the energy flows over $k_1$ is reduced, the battery level of each slot does not increase and no new $B^{\max}$-violation point emerges in $[a+1,c]$.

For a consecutive recursion backward search, we recursively perform the backward search on $[a+1,k_i]$ till the transmission schedule of $[a+1,k_i]$ is semi-feasible, or add $k_i$ (also denoted as $b$) as BFP if the transmission schedule of $[a+1,k_1]$ (also denoted as $[a+1,b]$) is feasible. Following this process, before $b$ is added as BFP, we obtain a series of BFPs $k_i$ such that the water level of $[a+1,b]$ is higher than that of $[b+1,k_{l-1}],[k_{l-1}+1,k_{l-2}],\ldots,[k_1+1,c]$, e.g., $k_1,k_2,k_3$ in Fig.~\ref{fg:l3_1}. Specifically, the transmission schedule of $[a+1,b]$ is feasible while the transmission schedule for all other segments separated by the newly set BFPs may not be feasible with only violation point $m$ such that $b_m<0$.

After the above consecutive recursion steps of the backward search, $[a+1,c]$ is separated by a set of BFP $k_i$ and we have the water level corresponding to each subsegment, e.g., $w_j$ in Fig.~\ref{fg:l3_1}. Then, we want to partially change the water levels to get the BFPs $b_i$ and BDP $k$ which we need, e.g., $b_1,k$ in Fig.~\ref{fg:l3_2}. To generate the series of $b_i$ and $k$, we can first check if $w_2$ makes the transmission schedule of the $w_2$-corresponding segment feasible. If not, we perform the energy push operation on the $w_2$-corresponding segment: gradually lower the water level before the first violation point $m$ such that $B^{m}<0$ to push more energy flow to $m$ until it becomes BDP. Specifically, when we gradually lower the water level, if the battery becomes full in some slots, we keep them as BFP, stop lowering the water level before the new BFP but continue gradually lowering the water level after the BFP, until $m$ becomes BDP. Obviously, the new emerged BFPs and BDP can be considered as $b_i$ and $k$ which we want to find.

On the other hand, if the transmission schedule of the $w_2$-corresponding segment is feasible, as illustrated in Fig.~\ref{fg:l3_2}, we then compare $w_3$ and $w_4$. If $v_3 > v_4$, we repeat the above procedure on $v_3$-corresponding segment; otherwise, as illustrated in Fig.~\ref{fg:l3_1}, $v_3>v_4$, we merge $w_3$-corresponding segment and $w_4$-corresponding segment to be a new segment, e.g., $[k_3+1,k_1]$ in Fig.~\ref{fg:l3_1},  and then get the water level by \eqref{eq:swfc}. Specifically, the water level of the newly merged segment is lower than $w_2$ since $w_2>w_4>w_3$. For the new merged segment, we repeat the above procedures. For the example in Fig.~\ref{fg:l3_2}, the transmission schedule of the merged segment is in-feasible, then $k$ is the BDP obtained by the energy pushing operation.


Obviously, the above iterative procedures can only stop when $k$ is found (or the water level of $[b_l+1,c]$ is feasible, i.e., we can consider $c$ as $k$). Following the above procedure, for all segments bounded by BFPs and ending at $k$, the water levels can only decrease segment by segment, i.e., we find the required BFPs $b_i$ and BDP $k$.

\section{Proof of Theorem 4}
Necessity: Since the schedule given by $\cal P$ and ${\cal D}^*$ is the optimal solution to the problem in \eqref{eq:problem}-\eqref{eq:cst}, the K.K.T. conditions are satisfied. Since the K.K.T. conditions of  \eqref{eq:osp} are a subset of those of the problem in \eqref{eq:problem}-\eqref{eq:cst}, $\boldsymbol{p}_n\in{\cal P}$ is the optimal solution to \eqref{eq:osp} for each transmitter $n$. Sufficiency: Given ${\cal D}^*$, since $\boldsymbol{p}_n$ is the optimal solution to \eqref{eq:osp}, the K.K.T. conditions are satisfied. Note that the K.K.T. conditions of  \eqref{eq:osp} are a subset of those of \eqref{eq:problem}-\eqref{eq:cst}. If all transmitters, $n\in{\cal N}$, satisfy the above conditions at the same time, all KKT conditions of the problem in \eqref{eq:problem}-\eqref{eq:cst} are satisfied by $\cal P$, i.e., $\cal P$ is the optimal solution to the problem in \eqref{eq:problem}-\eqref{eq:cst}.

\bibliographystyle{IEEETran}
\bibliography{IEEEabrv,bib}

\end{document}